\documentstyle[11pt,fleqn]{article}
\date{}
\newcommand{\eeq}{\end{eqnarray}}
\newcommand{\beq}{\begin{eqnarray}}
\newcommand{\bD}{{\bf D}}

\newcommand{\bB}{{\bf B}}

\newcommand{\bR}{{\bf R}}

\newcommand{\cL}{{\cal L}}
\newcommand{\cJ}{{\cal J}}

\newcommand{\cH}{{\cal H}}

\newcommand{\cD}{{\cal D}}
\newcommand{\cG}{{\cal G}}
\newcommand{\cB}{{\cal B}}
\newcommand{\vp}{{\varphi}}

\newcommand{\cP}{{\cal P}}
\newcommand{\p}{{\partial}}

\newcommand{\nb}{{\nabla}}

\def\con{{}_{\_\rule{-1pt}{0pt}\_}
\rule{-2pt}{0pt}\raise1.5pt\hbox{$\mid$}\hspace{2pt}}
\def\theequation{\arabic{section}.\arabic{equation}}

\title{\bf Quasi-local structure of  $p$-form theory}
\author{Dariusz Chru\'sci\'nski\footnotemark \\
 Institute of Physics, Nicholas Copernicus University\\
 ul. Grudzi\c{a}dzka 5/7, 87-100 Toru\'n, Poland}

\begin{document}
\def\thefootnote{\relax}\footnotetext{$^*$E-mail:
darch@phys.uni.torun.pl}

\maketitle

\begin{abstract}

We show that the Hamiltonian dynamics of the self-interacting, abelian
$p$-form theory in $D=2p+2$ dimensional space-time gives rise to the {\it
quasi-local} structure. Roughly
speaking, it means that  the field energy is localized but on closed
$2p$-dimensional surfaces (quasi-localised).
From the mathematical point of view this approach
is implied by the boundary value problem for the corresponding field
equations. Various boundary problems, e.g. Dirichlet or Neumann, lead to
different Hamiltonian dynamics. Physics seems to prefer gauge-invariant,
positively defined Hamiltonians which turn out to be quasi-local.
Our approach is closely related with the standard two-potential
formulation and enables one to generate e.g. duality transformations
in a perfectly local way (but with respect to a new set of nonlocal
variables). Moreover, the form of the quantization condition
displays very similar structure to that of the symplectic form of the
underlying $p$-form theory expressed in the quasi-local language.

\vspace{.5cm}

Keywords: p-form theory, duality invariance, Hamiltonian dynamics

PACS numbers: 11.15-q, 11.10Kk, 10.10Lm

\end{abstract}

\section{Introduction}
\setcounter{equation}{0}

One of the most important idea of modern physics is {\em locality}. It is
strongly related with relativity and quantum mechanics and plays a central
role in  relativistic (classical and quantum) field theories. Let us cite
only two very influential books: {\em physics is simple when analyzed
locally} \cite{MTW} and {\em the role of fields is to implement the
principle of locality} \cite{Haag}. It should be stressed, therefore, from
the very beginning that  we are not going to discuss nonlocal theories.
The abelian $p$-form theory is a simple generalization of an ordinary
electrodynamics in 4-dimensional Minkowski space-time ${\cal M}^4$
where the electromagnetic field potential 1-form $A_\mu$ is replaced by a
$p$-form in $D$-dimensional space-time \cite{Teitel1}, \cite{Nep}.
This theory is perfectly local, i.e. it is defined {\em via} the local
Lagrangian.

The motivations to study $p$-form theory are already discussed
in \cite{Teitel1}. Recently the new input comes with electric-magnetic
duality  \cite{Gibbons}, \cite{Deser1}, \cite{Deser2}.
 It was observed long
ago \cite{Deser-old} that the duality symmetry for the standard Maxwell
electrodynamics in four dimensional Minkowski space-time (i.e. $p=1$
theory) is generated by the nonlocal generator (its physical interpretation
as a chirality operator was discussed in \cite{IBB}), i.e. it is nonlocal
functional of the electromagnetic field. Therefore, the nonlocality enters
into the game in a very natural way. We shall see that the above mentioned
nonlocality  is closely related with the Hamiltonian
description of the field dynamics.

To define the Hamiltonian evolution one
splits the entire space-time into space and time and then formulates the
initial value problem. But in field theory one has to specify also the
boundary condition for the fields. Very often one assumes that all fields
do vanish at spatial infinity and simply forgets about this problem. It
should be stressed, however,  that even if the boundary values vanish
numerically
they do not vanish functionally, i.e. they are necessary in the proper
definition of the functional phase space of the dynamical problem. This is
typical for the problems with infinitely many degrees of freedom. Boundary
value problem is not only a mathematical problem. It also does belong to
physics.
Different boundary problems lead to different Hamiltonians, i.e. different
definitions of the field energy, e.g. energies defined {\it via} canonical
and symmetric energy-momentum tensors. Now, in the standard (i.e. $p=1$)
electrodynamics  the ``canonical'' energy, which is neither
gauge-invariant nor positively defined, is related to the
boundary value problem for the scalar potential $A_0$. On the other hand
the ``symmetric energy'' (defined by the symmetric energy-momentum
tensor), which is perfectly gauge-invariant and positively defined, is
related to the control of the electric and magnetic fluxes on the
boundary
 \cite{Jacek}, \cite{Kij-Dar}, \cite{Kijowski}. \cite{HAM}.
Therefore, it distinguishes a new set of electromagnetical variables
$Q^1$ and $Q^2$ consistent with the boundary problem. Together with the
canonically conjugated momenta $\Pi_1$ and $\Pi_2$ they encode the entire
gauge-invariant information about the electromagnetic field $F=dA$, i.e.
knowing $Q$'s and $\Pi$'s one may uniquely reconstruct $F$ \cite{Jacek}.
Actually, it was shown long ago by Debye \cite{Debye} that Maxwell theory
could be described in terms of two complex functions (so called Debey
potentials). It turns out that this formulation is very well suited to
describe e.g. radiative phenomena \cite{rad}. Our $Q$'s and $\Pi$'s (they may
be rearranged into  complex $Q$ and $\Pi$) are closely related to Debey
potentials. They solve the Gauss
constraint and, therefore, they reduce the symplectic form in the space of
 Cauchy data for the field dynamics.
However, they are nonlocal functions of the electromagnetic fields $\bD$
and $\bB$. The
nonlocality is of the very special structure and the Hamiltonian
generating the dynamics defines
a {\it quasi-local} functional, i.e. performing an integration over angle
variables one obtains perfectly local functional.

Now, in the abelian self-interacting $p$-form theory in $D=2p+2$
dimensional space-time one may perform the
similar analysis \cite{ROMP}:
 instead of two complex functions $Q$ and $\Pi$, the dynamical information
about a $p$-form electromagnetic fields $D$ and $B$  is now encoded into
two complex
$(p-1)$--forms.
In the present paper we relate the quasi-local picture implied by these
$(p-1)$--forms  with
the proper definition of the Hamiltonian dynamics for a $p$-form theory.
Moreover, we show that this formulation is perfectly suited for the
description of the duality symmetry, i.e. the duality rotations (for odd $p$)
are generated locally in terms of $Q$ and $\Pi$. We show that the canonical
generator has the following form:
\beq            \label{CG1}
\int Q^1\Pi_2 - Q^2\Pi_1\ .
\eeq
 It is evident that this
approach is closely related to the two-potential formulation
\cite{Deser2}, \cite{SS} (see Appendix~D).

It is well known that there is a crucial difference between theories with
different parities of $p$, e.g. for even $p$ a theory can not be duality
invariant. Now, it was observed only recently \cite{Deser2} that the
quantization
condition for $(p-1)$--brane dyons crucially depends upon $p$, namely
\begin{equation}  \label{QC}
e_1 g_2 + (-1)^p e_2 g_1 = nh\ ,
\end{equation}
with integer $n$ ($h$ is the Planck constant). It turns out that the
symplectic form of a $p$-form theory written in terms of $Q$ and $\Pi$
has very  similar structure
\begin{equation}    \label{Omega}
\Omega_p = \int \delta \Pi_1 \wedge \delta Q^1 + (-1)^{p+1}
\delta \Pi_2 \wedge \delta Q^2\ ,
\end{equation}
therefore, there is a striking correspondence between the form of the
quantization condition (\ref{QC}) and the structure of symplectic form
(\ref{Omega}).
 This correspondence is universal, i.e. it holds for any
gauge-invariant, self-interacting theory.

The paper is organized as follows: we remind the quasi-local structure of
standard (1-form) electrodynamics  in Section~2. This is the prototype of
the $p$-form theory for odd $p$.
Then in Section~3 we make
the generalization for $p=2$ which is the prototype for even $p$. The
general case (i.e. an arbitrary $p$) is
discussed in Appendices~B and C. In Section~4 we describe the gauge-invariant
coupling of
$p$-form
electrodynamics to the charged matter and the Hamiltonian structure of
the interacting theory.
The details of notation are clarified in Appendix~A.

\section{1-form theory in $D=4$}    \label{1FORM}
\setcounter{equation}{0}

\subsection{Generating formula}

Let us consider a $1$-form theory defined by the Lagrangian ${\cal L} =
{\cal L}(A,\partial A)$. Field dynamics of this theory may be written in
terms of the following generating formula (see Appendix~A for details of
notation):
\beq   \label{gen1}
- \delta \cL =  \partial_\nu (\cG^{\nu\mu}\delta
A_{\mu}) =  (\partial_\nu \cG^{\nu\mu})\delta
A_{\mu} +  \cG^{\nu\mu} \delta (\partial_\nu
A_{\mu}) \ .
\eeq
The formula (\ref{gen1}) implies the following definition of ``momenta'':
\beq   \label{cG1}
\cG^{\mu\nu} =
 - 2\,\frac{\partial \cL}{\partial
F_{\mu\nu}}\ .
\eeq
Moreover,  (\ref{gen1}) generates  dynamical (in general nonlinear) field
equations
\beq   \label{f-eqs}
\partial_\nu \cG^{\nu\mu} = - \cJ^{\mu}\ ,
\eeq
where the external $1$-form current reads:
\beq            \label{J}
 \cJ^{\mu} =
  \frac{\partial \cL}{\partial A_{\mu}}\ .
\eeq
Let us start with a source  free theory, i.e. $\cJ=0$. We shall study the
$p$-form electromagnetism coupled to a charged matter in section~\ref{MATTER}.
 To obtain the Hamiltonian description of the field dynamics
let us integrate equation (\ref{gen1}) over a 3-dimensional volume $V$
contained in the constant-time hyperplane $\Sigma$:
\beq   \label{1-1}
- \delta\ \int_{V} \cL = \int_{V} \partial_0( \cG^{0i}\delta
A_{i}) + \int_{\partial V} \cG^{\perp\mu}\delta
A_{\mu}\ ,
\eeq
where $\perp$ denotes the component orthogonal to the 2-dimensional
boundary $\p V$.
 To simplify our
notation let us introduce the spherical coordinates on $\Sigma$:
\beq   \label{s1}
x^{3} = r\ , \ \ \ \ x^A = \vp_A\ ;\ \ \ \
A=1,2\ , \eeq where $\vp_1,\vp_2$ denote
spherical angles (usually one writes $\vp_1=\vp$ and $\vp_2 = \Theta$). To
enumerate angles we shall use capital letters
$A,B,C,...$. The Euclidean metric tensor is diagonal
\beq
g_{11} = r^2, \ \ \ g_{22} = r^2\sin\vp_2, \ \ \ g_{rr}=1\ ,
\eeq
and the volume form $\Lambda_1 = \sqrt{\det(g_{kl})} = r^2\sin\vp_2$.
Let $V$ be
 a 3-ball with a finite radius $R$. In such a coordinate system the
formula
(\ref{1-1}) takes the following form:
\beq  \label{2-1}
\delta\ \int_{V} \cL &=&  - \int_{V} \partial_0( \cD^{i}
\delta A_{i}) + \int_{\partial V} \cD^{r}\delta
A_{0}         -
\int_{\partial V} \cG^{rB}\delta A_{B}\ ,
\eeq
where
\begin{equation}
\cD_{i} = \cG_{i0}
\end{equation}
denotes the $1$-form electric induction density on $\Sigma$.
Now, performing the Legendre transformation between induction $1$-form
$\cD^{i}$ and $\dot{A}_{i}$ one obtains the following
Hamiltonian formula:
\beq   \label{3-1}
- \delta \cH_{can} &=& - \int_{V} \left( \dot{{\cD}}^{i} \delta
A_{i} - \dot{A}_{i} \delta \cD^{i} \right)
+ \int_{\partial V}  \cD^{r}\delta A_{0}
- \int_{\partial V} \cG^{rB}\delta A_{B}\ ,
\eeq
where the canonical Hamiltonian
\beq   \label{HP-1}
\cH_{can} = \int_{V} \left( - \cD^{i}\dot{A}_{i} - \cL
\right)\ .
\eeq
Equation (\ref{3-1}) generates an infinite-dimensional Hamiltonian system in
the phase space ${\cal P}_p = (\cD^{i},A_{i})$
fulfilling Dirichlet boundary conditions for the $1$-form potential
$A_{i}$: $A_{0}|\partial V$ and $A_{A}|\partial
V$. From the mathematical point of view this is the missing part of the
definition of the functional space. The Hamiltonian structure of a general
nonlinear $1$-form electrodynamics described above is mathematically well
defined, i.e. a mixed Cauchy problem (Cauchy data given on $\Sigma$ and
Dirichlet data given on $\p V \times \bR$) has a unique solution (modulo gauge
transformations which reduce to the identity on $\p V \times \bR$).

There is, however, another way to describe the Hamiltonian evolution of
fields in the region $V$. Let us perform the Legendre transformation
between $\cD^{r}$ and $A_{0}$ at the boundary $\partial
V$. One obtains:
\beq   \label{4-1}
- \delta {\cH}_{sym} &=& - \int_{V} \left(
\dot{{\cD}}^{i}
\delta
A_{i} - \dot{A}_{i} \delta \cD^{i} \right)
- \int_{\partial V} A_{0}\delta \cD^{r}
- \int_{\partial V} \cG^{rB}\delta A_{B}\ ,
\eeq
where the new ``symmetric'' Hamiltonian
\beq   \label{Hpo-1}
{{\cH}}_{sym} = \cH_{can}  + \int_{\partial V} \cD^{r}
{A}_{0} \ .
\eeq
Observe, that formula (\ref{4-1}) defines the Hamiltonian evolution but on a
different phase space. In order to kill boundary terms in (\ref{4-1}) one
has to
control on $\partial V$: $\cD^{r}$ (instead of $A_{0}$) and
$A_{B}$. We stress that from the mathematical point of view both
descriptions are equally good and an additional physical argument has to be
given if we want to choose one of them as more fundamental.

\subsection{Canonical vs. symmetric energy}

Now, let us discuss the relation between $\cH_{can}$ and ${{\cH}}_{sym}$
defined by (\ref{HP-1}) and (\ref{Hpo-1}) respectively. One has:
\beq   \label{HPo1}
{{\cH}}_{sym} &=& \cH_{can}  + \int_{\partial V} \cD^{r}
{A}_{0}
= \cH + \int_{V} \partial_k\left( \cD^{k}
{A}_{0} \right) \nonumber\\
&=& \int_{V} \left\{ - \cD^{i}\dot{A}_{i} - \cL
 +  \left(
{A}_{0}\partial_k\cD^{k} + \cD^{k}
\partial_k{A}_{0} \right) \right\}    \nonumber\\
&=& \int_{V} \left( \cD^{i}E_{i} -
\cL\right)\ ,
\eeq
where the  $1$-form electric field is defined by
\begin{equation}
E_{i} = F_{i0} = \p_{[i}A_{0]}\ .
\end{equation}

Therefore, $\cH_{sym}$ is related to $\cL$ {\em via}
different Legendre  transformation (compare (\ref{HP-1}) with
(\ref{HPo1})). Contrary to $\cH_{can}$, ${{\cH}}_{sym}$ is
perfectly gauge-invariant. It is evident that ${{\cH}}_{sym}$
is defined {\it via} the symmetric energy-momentum tensor:
\begin{equation}  \label{Ts-1}
T^{\mu\nu}_{sym} =  F^{\mu\lambda}\cG^\nu_{\ \lambda} +
g^{\mu\nu}\cL\ ,
\end{equation}
whereas $\cH_{can}$ {\it via} the canonical one:
\begin{equation}  \label{Tc-1}
T^{\mu\nu}_{can} = (\p^\mu A^{\lambda}) G^{\nu}_{\ \lambda}
+ g^{\mu\nu}\cL\ ,
\end{equation}
i.e. $\cH_{sym} = \int_{V} T^{00}_{sym}$ and ${\cH}_{can} =
\int_{V} T^{00}_{can}$. Therefore, the ``symmetric energy''
$\cH_{sym}$ is gauge-invariant and positively defined,
e.g. for the $1$-form Maxwell theory one has
\beq
\cH^{Maxwell}_{sym}
= \frac{1}{2} \int (\cD^{i}D_{i} +
\cB^{i}B_{i} )\ .\nonumber
\eeq
 On the other hand, the ``canonical energy'' $\cH_{can}$ is neither
positively defined nor gauge-invariant.
These properties show that the Hamiltonian evolution based on
$\cH_{sym}$ is more natural from  the physical point of view than the
one  based on $\cH_{can}$ (see also discussion in \cite{Kijowski}).

\subsection{Reduction of the generating formula}

Now, it turns out that the formula (\ref{4-1}) may be considerably
simplified.
Any geometrical object on a 3-dimensional hyperplane $\Sigma$ may be
decomposed into the radial and tangential (i.e. tangential to any sphere
$S^2(r)$) components, e.g. a $1$-form gauge
potential $A_{i}$ decomposes into the radial $A_{r}$ and
tangential $A_{A}$.  Now, any $1$-form on $S^{2}(r)$ may be further
decomposed into
``longitudinal'' and ``transversal'' parts:
\begin{equation}   \label{dec-1}
A_{A} = \nabla_{A}u  +
\epsilon_{AB}\nabla^{B} v\ ,
\end{equation}
where both $u$ and $v$ are scalar functions on $S^{2}(r)$. Now, using
(\ref{dec-1}) and
integrating by parts one gets:
\beq  \label{4a-1}
\lefteqn{ \int_{V} \left(
\dot{{\cD}}^{i}
\delta
A_{i} - \dot{A}_{i} \delta \cD^{i}
\right)        =
\int_{V}  \left\{ \left(
\dot{{\cD}}^{r}
\delta
A_{r} - \dot{A}_{r} \delta \cD^{r} \right) \right. }  \nonumber\\
    &+& \left.
  \left[
(\p_r\dot{{\cD}}^{r})
\delta
u - \dot{u} \delta (\p_r\cD^{r}) \right] -
\epsilon_{AB} \left[
\left(\nb^{B}\dot{{\cD}}^{A}\right)
\delta
v - \dot{v} \delta \left(\nb^{B}\cD^{A}\right) \right]   \right\}  \ ,
\eeq
where we have used the Gauss law
\begin{equation}
\nb_{A} \cD^{A} = - \p_r \cD^{r}\ .
\end{equation}
Moreover, due to (\ref{dec-1})
\beq    \label{5-1}
\int_{\p V} {\cG}^{rA} \delta A_{A} =
 - \int_{\p V} \left\{ - \dot{{\cD}}^r \delta u
+ \left( \epsilon_{AB} \nb^{B} \cG^{rA} \right) \delta
v\right\}\ .
\eeq
In deriving (\ref{5-1}) we have used
\begin{equation}
\nb_{A} \cG^{Ar} = - \dot{{\cD}}^{r}\ ,
\end{equation}
which follows from the field equations $\nb_{A} \cG^{Ar} + \p_0
\cG^{0r} = 0$.
Now, taking into account (\ref{4a-1}) and (\ref{5-1}) the generating formula
(\ref{4-1}) may be rewritten in the following way:
\beq   \label{6-1}
- \delta \cH_{sym} &=& - \int_{V} \left\{ \left[
\dot{{\cD}}^{r} \delta \left( A_{r} - \p_r u \right) -
\left( \dot{A}_{r} - \p_r \dot{u} \right) \delta
\cD^{r} \right] \right. \nonumber\\
&-& \left.  \left[ \left(\epsilon_{AB} \nb^{B}
\dot{{\cD}}^{A}\right) \delta v -
 \dot{v} \delta\left(\epsilon_{AB}
\nb^{B} \cD^{A} \right) \right]  \right\} \nonumber \\
&-&  \int_{\p V} \left\{ \left( A_{0} -  \dot{u}\right) \delta
\cD^{r} -
 \left(\epsilon_{AB} \nb^{B} \cG^{rA} \right)
\delta v \right\}  \ .
\eeq
Note, that although $A_{r}$, $A_{0}$ and $u$ are
manifestly gauge-dependent, the combinations
$ A_{r} -  \p_r u$ and
$ A_{0} -  \p_0 {u}$
are gauge-invariant. To simplify our consideration we choose the special gauge
$u\equiv 0$, i.e. a $1$-form $A_{A}$ on $S^{2}(r)$ is purely
transversal. This condition, due to (\ref{dec-1}), may be equivalently rewritten
as
\begin{equation} \label{gauge1-1}
\nb_{A} A^{A} =0\ .
\end{equation}
Assuming (\ref{gauge1-1}) one may show \cite{ROMP}
\beq
\Delta_{0}\, A^{r} =  {r^2}\,
\epsilon^{AB}\, \nb_{B} B_{A}\ ,
\eeq
where
\beq   \label{Delta-1}
\Delta_0 = r^2 \, \nb_A \nb^A
\eeq
denotes the 2-dimensional Laplacian on $S^2(1)$, i.e. the 2-dim.
Laplace-Beltrami operator on scalar functions ($0$-forms). Moreover,
\begin{equation}   \label{Br}
B^r = \epsilon^{AB}\, \nb_A A_B = - r^{-2} \Delta_0 v\ .
\end{equation}
Since $\Delta_0$ is
invertible in the source free theory \cite{Jacek} the formula (\ref{6-1})
may be
rewritten as follows:
\beq   \label{7-1}
- \delta \cH_{sym} &=& - \int_{V} \left\{ \left[
(r\dot{{\cD}}^{r}) \delta \left( r\Delta_0^{-1}\epsilon_{AB}\, \nb^B B^A
\right)
- \left( r\Delta_0^{-1}\epsilon_{AB}\, \nb^B \dot{B}^A \right)   \delta
(r\cD^{r}) \right] \right. \nonumber\\
&+& \left.  \left[ \left(r\Delta_0^{-1}\epsilon_{AB} \nb^{B}
\dot{{\cD}}^{A}\right) \delta (rB^r) -
 (r\dot{B}^r) \delta\left(r\Delta_0^{-1}\epsilon_{AB}
\nb^{B} \cD^{A}\right) \right]  \right\} \nonumber \\
&-&  \int_{\p V} \left\{ (r^{-1} A_{0})  \delta
(r \cD^{r}) +
 \left(\Delta_0^{-1}\epsilon_{AB} \nb^{B} \cG^{rA} \right)
\delta (rB^r) \right\}  \ .
\eeq
Now, introducing the following set of variables
\beq
Q^1 &=& rD^r\ ,\\
Q^2 &=& rB^r\ ,\\
\Pi_1 &=&  r \Delta_0^{-1}\epsilon^{AB} \nb_{B} B_A\ ,\\
\Pi_2 &=& -  r \Delta_0^{-1}\epsilon^{AB} \nb_{B} D_A\ ,
\eeq
eq. (\ref{7-1}) simplifies to
\beq  \label{8-1}
- \delta \cH_{sym} &=&  \int_{V}\Lambda_1 \left\{ \left( \dot{{\Pi}}^1
\delta Q_1 - \dot{Q}_1 \delta \Pi^1 \right)  +
\left( \dot{{\Pi}}^2
\delta Q_2 - \dot{Q}_2 \delta \Pi^2 \right)
\right\} \nonumber \\ &+&
 \int_{\p V}\Lambda_1 \left(\chi^1 \delta Q_1 + \chi^2 \delta Q_2 \right)\ ,
\eeq
where we introduced the boundary momenta:
\beq
\chi_1 &=& - \frac{1}{r} A_0\ ,\\
\chi_2 &=& - r \Delta_0^{-1}\epsilon_{AB} \nb^{B} G^{rA}\ .
\eeq
Tensor $G^{\mu\nu}$ is defined by $\cG^{\mu\nu} = \Lambda_1\, G^{\mu\nu}$,
and, therefore, $\cD^i = \Lambda_1\, D^i$.
Note, that
\beq   \label{chi-1}
\chi^l = \frac{\delta \cH_{sym}}{\delta (\p_r Q_l)}\ ,  \ \ \ \ \ \
l=1,2\ .
\eeq
For a Maxwell theory one obtains
\beq
\cH_{sym}^{Maxwell} &=& \frac 12 \int_V \Lambda_1 \sum_{l=1}^2
\left\{  \frac{1}{r^2} Q_l Q_l -
\frac{1}{r^2} \p_r (rQ_l) \Delta_0^{-1} \p_r(rQ_l) - \Pi^l \Delta_0 \Pi^l
\right\}\ ,
\eeq
and, therefore
\beq    \label{chi-11}
\chi^l = \frac{1}{r}\,\Delta_0^{-1} \p_r(rQ_l)\ , \ \ \ \ \ \ l=1,2\ ,
\eeq
have perfectly symmetric form.

\subsection{Canonical symmetries}

The symplectic form $\int \delta \cD^k \wedge \delta A_k$ rewritten in terms
of
$Q$'s and $\Pi$'s have the following form \cite{Jacek}, \cite{ROMP}:
\begin{equation}  \label{Om-1}
\Omega = \mbox{Im}\, \int \Lambda_1 \, \delta\Pi \wedge \delta \overline{Q}\ ,
\end{equation}
where we introduced a complex notation
\beq
Q &=& Q^1 +iQ^2\ , \\
\Pi &=& i(\Pi_1 + i\Pi_2)\ .
\eeq
The form (\ref{Om-1}) is invariant under the following set of
$\bR$-linear transformations:
\beq        \label{1S}
Q &\rightarrow& e^{i\alpha} \, Q\ ,\\   \label{2S}
Q &\rightarrow& \cosh\alpha\, Q + i\sinh\alpha\, \overline{Q}\ ,\\  \label{3S}
Q &\rightarrow& \cosh\lambda\, Q + \sinh\lambda\, \overline{Q}\ ,
\eeq
and the same rules for $\Pi$. It is easy to see that these transformations
form the group $SO(2,1)$. In terms of $\bD$ and $\bB$, (\ref{1S})--(\ref{3S})
have more familiar form: \\
(\ref{1S}) corresponds to orthogonal $SO(2)$ duality rotations:
\beq   \label{I}
\bD &\rightarrow & \bD \cos\alpha - \bB\sin\alpha\ ,\nonumber\\
\bB &\rightarrow & \bD \sin\alpha + \bB\cos\alpha\ ,
\eeq
(\ref{2S}) corresponds to hyperbolic $SO(1,1)$ rotations:
\beq         \label{II}
\bD &\rightarrow & \bD \cosh\alpha + \bB\sinh\alpha\ ,\nonumber\\
\bB &\rightarrow & \bD \sinh\alpha + \bB\cosh\alpha\ ,
\eeq
(\ref{3S}) corresponds to scaling transformations:
\beq                   \label{III}
\bD &\rightarrow & e^{\lambda}\,\bD \ ,\nonumber\\
\bB &\rightarrow & e^{-\lambda}\, \bB\ .
\eeq
The canonical generators corresponding to (\ref{1S})--(\ref{3S}) have the
following form:
\beq   \label{G1-1}
G_1 &=& \int \Lambda_1 \,(Q^2\Pi_1 - Q^1\Pi_2) = \mbox{Re}\, \int \Lambda_1\,
(\Pi\overline{Q})\ , \\   \label{G2-1}
G_2 &=& - \int \Lambda_1 \,(Q^2\Pi_1 + Q^1\Pi_2) = \mbox{Re}\, \int
\Lambda_1\,
(\Pi Q)\ , \\ \label{G3-1}
G_3 &=&  \int \Lambda_1 \,(Q^1\Pi_1 - Q^2\Pi_2) = \mbox{Im}\, \int \Lambda_1\,
(\Pi Q)\ .
\eeq
Note, that for the duality invariant theory $G_1$ defined in (\ref{G1-1}) is
constant in time. Its physical interpretation was clarified in \cite{IBB}.
Obviously, $G_1$, $G_2$ and $G_3$ rewritten in terms of $\bD$ and $\bB$ are
highly nonlocal functionals of the fields
\cite{IBB}, \cite{Deser-old}.

\subsection{Summary}

The reduced variables $(Q_l,\Pi^l)$ play the role of generalized positions
and momenta for an electromagnetic field. They are perfectly
gauge-invariant and contain the entire (gauge-invariant) information about
$\bD$ and $\bB$. Let us note that $Q$'s and $\Pi$'s are nonlocal functions of
$\bD$ and $\bB$. The nonlocality enters {\it via} the operations on each
sphere $S^2(r)$, i.e. {\it via} the operator
$\Delta^{-1}_0$. On the other hand the operations in the radial direction do
not produce any nonlocality.

The Hamiltonian generating the dynamics is perfectly local in $\bD$ and
$\bB$ but is nonlocal in $Q$'s and $\Pi$'s. The field functional  with the
above described nonlocality we shall call {\it quasi-local}.
Note, that generators $G_i$ are perfectly local in reduced variables.

The ``symmetric'' Hamiltonian dynamics is defined by the
Dirichlet boundary conditions for positions $Q_l$. On the other hand the
``canonical'' formula (\ref{4-1}) is defined by the Dirichlet boundary
condition for $\chi^1$ and $Q_2$. Note, however, that in the Maxwell case
\beq
\int_{\p V} \Lambda_1 Q_1 \delta \chi^1 =
\int_{\p V} \Lambda_1 \frac{1}{r} \left(\Delta_0^{-1} Q_1
\right)\delta\ \p_r(r^2D^r) =\int_{\p V} r \left(\Delta_0^{-1} Q_1
\right)\delta\  (\p_r\cD^r)\ ,
\eeq
i.e. a Dirichlet condition $\chi^1|\p V$ is equivalent to the Neumann
condition $\p_r \cD^r|\p V$.

\section{2-form theory in $D=6$}
\setcounter{equation}{0}

\subsection{Generating formula}

Now,  consider a $2$-form theory defined by the Lagrangian ${\cal L} =
{\cal L}(A,\partial A)$. Field dynamics of this theory may be written in
terms of the following generating formula:
\beq   \label{gen-2}
- \delta \cL =  \partial_\nu (\cG^{\nu\mu\lambda}\delta
A_{\mu\lambda}) =  (\partial_\nu \cG^{\nu\mu\lambda})\delta
A_{\mu\lambda} +  \cG^{\nu\mu\lambda} \delta (\partial_\nu
A_{\mu\lambda}) \ .
\eeq
The formula (\ref{gen-2}) implies the following definition of ``momenta'':
\beq   \label{cG}
\cG^{\mu\nu\lambda} =
 - 3!\,\frac{\partial \cL}{\partial
F_{\mu\nu\lambda}}\ .
\eeq
Moreover,  (\ref{gen-2}) generates  dynamical (in general nonlinear) field
equations
\beq   \label{f-eqs2}
\partial_\nu \cG^{\nu\mu\lambda} = - \cJ^{\mu\lambda}\ ,
\eeq
where the external $2$-form current reads:
\beq            \label{J-2}
 \cJ^{\mu\lambda} =  2\,
  \frac{\partial \cL}{\partial A_{\mu\lambda}}\ .
\eeq
In the presents section we consider only $\cJ=0$ (for $\cJ \neq 0$ see
section~\ref{MATTER}.)
 To obtain the Hamiltonian description of the field dynamics
let us integrate equation (\ref{gen-2}) over a 5-dimensional volume $V$
contained in the constant-time hyperplane $\Sigma$:
\beq   \label{1-2}
- \delta\, \int_{V} \cL\  =\  \int_{V} \partial_0( \cG^{0ij}\delta
A_{ij}) + \int_{\partial V} \cG^{\perp\mu\nu}\delta
A_{\mu\nu}\ ,
\eeq
where $\perp$ denotes the component orthogonal to the $4$-dimensional
boundary $\p V$.
 To simplify our
notation let us introduce the spherical coordinates on $\Sigma$:
\beq   \label{s2}
x^{5} &=& r\ , \ \ \ \ x^A = \vp_A\ ;\ \ \ \
A=1,2,3,4\ , \eeq
where $\vp_1,\vp_2,\vp_{3},\vp_4$ denote
spherical angles (to enumerate angles we shall use capital letters
$A,B,C,...$). The Euclidean metric on $\Sigma$ reads:
\beq
\lefteqn{ g_{11} = r^2\sin^2\vp_2\sin^2\vp_3\sin^2\vp_4 \ ,\ \
 \ g_{22}=r^2\sin^2\vp_3\sin^2\vp_4 ,\ } \nonumber\\ && g_{33} =
r^2\sin^2\vp_4 \ ,\ \ \
 g_{44}= r^2\ ,\ \ \ g_{55}\equiv
g_{rr}=1\ ,
\eeq
and the corresponding volume form
\beq
\Lambda_2 = \sqrt{\det(g_{ij})} = r^4\sin\vp_2\sin^2\vp_3\sin^3\vp_4\ .
\eeq
Let $V$ be
 a 5-dim. ball with a finite radius $R$. In such a coordinate system the
formula
(\ref{1-2}) takes the following form:
\beq  \label{2-2}
\delta\ \int_{V} \cL &=&  \int_{V} \partial_0( \cD^{ij}
\delta A_{ij}) -  \int_{\partial V} 2\,\cD^{rA}\delta
A_{0A}          -
\int_{\partial V} \cG^{rAB}\delta A_{AB}\ ,
\eeq
where
\begin{equation}
\cD_{ij} = \cG_{ij0}
\end{equation}
denotes the $2$-form electric induction density.
Now, performing the Legendre transformation between induction $2$-form
$\cD^{ij}$ and $\dot{A}_{ij}$ one obtains the following
Hamiltonian formula:
\beq   \label{3-2}
- \delta \cH_{can} &=& \int_{V} \left( \dot{{\cD}}^{ij} \delta
A_{ij} - \dot{A}_{ij} \delta \cD^{ij} \right)
-  \int_{\partial V} 2\, \cD^{rA}\delta A_{0A}
- \int_{\partial V} \cG^{rAB}\delta A_{AB}\ ,
\eeq
where the canonical Hamiltonian
\beq   \label{Hp-2}
\cH_{can} = \int_{V} \left(  \cD^{ij}\dot{A}_{ij} - \cL
\right)\ .
\eeq
Equation (\ref{3-2}) generates an infinite-dimensional Hamiltonian system in
the phase space ${\cal P}_2 = (\cD^{ij},A_{ij})$
fulfilling Dirichlet boundary conditions for the $2$-form potential
$A_{ij}$: $A_{0A}|\partial V$ and $A_{AB}|\partial
V$. From the mathematical point of view this is the missing part of the
definition of the functional space. The Hamiltonian structure of a general
nonlinear $2$-form electrodynamics described above is mathematically well
defined, i.e. a mixed Cauchy problem (Cauchy data given on $\Sigma$ and
Dirichlet data given on $\p V \times \bR$) has a unique solution (modulo gauge
transformations which reduce to the identity on $\p V \times \bR$).

Note the difference in signs between corresponding formulae of the present
section and that of section \ref{1FORM}. This difference follows from the
difference between corresponding symplectic structures \cite{ROMP}. For
1-form theory one has
\begin{equation}   \label{Omega-1}
\Omega_1 = \int_{V} \delta \cG^{0i} \wedge \delta A_{i} = +
 \int_{V}\delta\cD^{i}\wedge \delta{A}_{i}\ ,
\end{equation}
whereas for 2-form theory
\begin{equation}   \label{Omega-2}
\Omega_2 = \int_{V} \delta \cG^{0ij} \wedge \delta A_{ij} = -
 \int_{V}\delta\cD^{ij}\wedge \delta{A}_{ij}\ ,
\end{equation}

Now, in analogy to (\ref{4-1}) we pass to another Hamiltonian description
of the field evolution  in the finite region $V$. Let us perform the Legendre transformation
between $\cD^{rA}$ and ${A}_{0A}$ at the boundary $\partial
V$. One obtains:
\beq   \label{4-2}
- \delta {{\cH}}_{sym} &=&  \int_{V} \left(
\dot{{\cD}}^{ij}
\delta
A_{ij} - \dot{A}_{ij} \delta \cD^{ij} \right)
+  \int_{\partial V} 2\,A_{0A}\delta \cD^{rA}
- \int_{\partial V} \cG^{rAB}\delta A_{AB}\ ,
\eeq
where the new ``symmetric'' Hamiltonian
\beq   \label{Hpo-2}
{{\cH}}_{sym} = \cH_{can}  -  \int_{\partial V}2\, \cD^{rA}
{A}_{0A} \ .
\eeq
Observe, that formula (\ref{4-2}) defines the Hamiltonian evolution but on a
different phase space. In order to kill boundary terms in (\ref{4-2}) one
has to
control on $\partial V$: $\cD^{rA}$ (instead of $A_{0A}$) and
$A_{AB}$. We stress that from the mathematical point of view both
descriptions are equally good and an additional physical argument has to be
given if we want to choose one of them as more fundamental.

\subsection{Canonical vs. symmetric energy}

The relation between $\cH_{can}$ and ${{\cH}}_{sym}$ is exactly the same
as in $p=1$ case:
\beq   \label{Hpo2}
{{\cH}}_{sym} &=& \cH_{can}  -  \int_{\partial V} 2\,\cD^{rA}
{A}_{0A}
= \cH_{can} -  \int_{V} 2\,\partial_k\left( \cD^{ki}
{A}_{0i} \right) \nonumber\\
&=& \int_{V} \left\{  \cD^{ij}\dot{A}_{ij} - \cL
+   2 \left(
{A}_{0i}\partial_k\cD^{ki} + \cD^{ki}
\partial_k{A}_{0i} \right) \right\}    \nonumber\\
&=& \int_{V} \left( \frac{1}{2}\,\cD^{ij}E_{ij} -
\cL\right)\ ,
\eeq
where the  $2$-form electric field is defined by
\begin{equation}
E_{ij} = F_{ij0} = \p_{[i}A_{j0]}\ .
\end{equation}
Therefore, $\cH_{sym} = \int T^{00}_{sym}$ and $\cH_{can} = \int
T^{00}_{can}$ with
\begin{equation}  \label{Ts-2}
T^{\mu\nu}_{sym} = \frac{1}{2} F^{\mu\lambda\sigma} G^\nu_{\ \lambda\sigma} +
g^{\mu\nu}\cL\ ,
\end{equation}
and
\begin{equation}  \label{Tc-2}
T^{\mu\nu}_{can} = (\p^\mu A^{\lambda\sigma}) G^{\nu}_{\ \lambda\sigma}
+ g^{\mu\nu}\cL\ .
\end{equation}
In the  $2$-form Maxwell theory the ``symmetric energy'' (gauge-invariant
and positively defined) reads:
\beq
\cH^{Maxwell}_{sym}
= \frac{1}{4} \int (\cD^{ij}D_{ij} +
\cB^{ij}B_{ij} )\ .\nonumber
\eeq

\subsection{Reduction of the generating formula}

Now, in analogy to (\ref{dec-1}) let as make the following decomposition:
\begin{equation}   \label{dec-2}
A_{AB} = \nabla_{[A}u_{B]} +
\epsilon_{ABCD}\nabla^{C} v^{D}\ ,
\end{equation}
where $\nb_A$ denotes a covariant derivative on each $S^4(r)$ defined by
the induced metric $g_{AB}$ and $\epsilon_{ABCD}$ stands for the
L\'evi-Civita tensor density such that $\epsilon_{1234} = \Lambda_2$.
Both $u_A$ and $v^A$ are $1$-forms on $S^{4}(r)$. Using (\ref{dec-2}) and
integrating by parts one gets:
\beq  \label{4a-2}
\lefteqn{\int_{V} \left(
\dot{{\cD}}^{ij}
\delta
A_{ij} - \dot{A}_{ij} \delta \cD^{ij}
\right)        =
\int_{V} \left\{ 2\, \left(
\dot{{\cD}}^{rA}
\delta
A_{rA} - \dot{A}_{rA} \delta \cD^{rA} \right) \right. } \nonumber\\
&+& \left.  2\, \left[
(\p_r\dot{{\cD}}^{rA})
\delta
u_{A} - \dot{u}_{A} \delta (\p_r\cD^{rA}) \right]
 - \epsilon_{ABCD} \left[
(\nb^{C}\dot{{\cD}}^{AB})
\delta
v^{D} - \dot{v}^{D} \delta (\nb^{C}\cD^{AB}) \right] \right\}
\ ,
\eeq
where we have used the Gauss law
\begin{equation}
\nb_{A} \cD^{AB} = - \p_r \cD^{rB}\ .
\end{equation}
Moreover, due to (\ref{dec-2})
\beq    \label{5-2}
\int_{\p V} {\cG}^{rAB} \delta A_{AB}  =
 - \int_{\p V} \left\{- 2\, \dot{\cD}^{rA} \delta u_A
+ \left(\epsilon_{ABCD} \nb^{C} \cG^{rAB}\right) \delta
v^{D}\right\}\ .
\eeq
In deriving (\ref{5-2}) we have used
\begin{equation}
\nb_{A} \cG^{ABr} = - \dot{{\cD}}^{rB}\ ,
\end{equation}
which follows from the field equations $\nb_{A} \cG^{ABr} + \p_0
\cG^{0Br} = 0$.
Now, taking into account (\ref{4a-2}) and (\ref{5-2}) the generating formula
(\ref{4-2}) may be rewritten in the following way:
\beq     \label{6-2}
- \delta \cH_{sym} &=& \int_{V} \left\{ \left[
\dot{{\cD}}^{rA} \delta ( 2\,A_{rA} - 2\p_r u_{A} )
- ( 2\dot{A}_{rA} - 2\p_r \dot{u}_{A} ) \delta
\cD^{rA} \right]
 \right.  \nonumber \\  &-&  \left.
\left[ (\epsilon_{ABCD} \nb^{C}
\dot{{\cD}}^{AB}) \delta v^{D}
-   \dot{v}^{D} \delta(\epsilon_{ABCD}
\nb^{C} \cD^{AB} \right]  \right\} \nonumber \\
&+&  \int_{\p V} \left\{ (2\, A_{0A} - 2\, \dot{u}_{A}) \delta
\cD^{rA}
+ \int_{\p V} (\epsilon_{ABCD} \nb^{C} \cG^{rAB} )
\delta v^{D} \right\} \ .
\eeq
Note, that although $A_{rA}$, $A_{0A}$ and $u_{A}$ are
manifestly gauge-dependent, the combinations
$ A_{rA} - \p_r u_{A}$ and
$ A_{0A} - \p_0 {u}_{A}$
are gauge-invariant. To simplify our consideration we choose the special gauge
$u\equiv 0$, i.e. a $2$-form $A_{AB}$ on $S^{4}(r)$ is purely
transversal. This condition, due to (\ref{dec-2}), may be equivalently rewritten
as
\begin{equation} \label{gauge1-2}
\nb_{A} A^{AB} =0\ .
\end{equation}
But now, contrary to the $p=1$ case, we have an additional covector field on
$S^4(r)$, namely $A_{rA}$. For this covector we choose an analogous gauge
condition, i.e.
\begin{equation} \label{gauge2-2}
\nb_{A} A^{rA} =0\ .
\end{equation}
Assuming (\ref{gauge1-2}) and (\ref{gauge2-2}) one may show \cite{ROMP}
\beq
\Delta_{1}\, A^{rD} = -  \frac{r^2}{4}\,
\epsilon^{ABCD}\, \nb_{C} B_{AB}\ ,
\eeq
where
\beq  \label{Delta-2}
\Delta_1 = r^2\nb_A \nb^A - 3\ ,
\eeq
 equals to the Laplace-Beltrami
operator on co-exact 1-forms on $S^4(1)$ \cite{ROMP}. Moreover, in analogy to
(\ref{Br}) one has \cite{ROMP}
\beq
B^{rA} = - 2 r^{-2} \Delta_1 v^A\ ,
\eeq
and, therefore, the formula (\ref{6-2}) simplifies to
\beq   \label{7-2}
- \delta \cH_{sym} &=& \frac 12  \int_{V} \left\{ - \left[
(r\dot{{\cD}}^{rD}) \delta \left( r\Delta_1^{-1}\epsilon_{ABCD}\,
\nb^C B^{AB}
\right)
- \left( r\Delta_1^{-1}\epsilon_{ABCD}\, \nb^C \dot{B}^{AB} \right)   \delta
(r\cD^{rD}) \right] \right. \nonumber\\
&+& \left.  \left[ \left(r\Delta_1^{-1}\epsilon_{ABCD} \nb^{C}
\dot{{\cD}}^{AB}\right) \delta (rB^{rD}) -
 (r\dot{B}^{rD}) \delta \left(r\Delta_1^{-1}\epsilon_{ABCD}
\nb^{C} \cD^{AB} \right) \right]  \right\} \nonumber \\
&+&  \int_{\p V} \left\{ (2r^{-1} A_{0A})  \delta
(r \cD^{rA}) - \left( \frac 12
r\Delta_1^{-1}\epsilon_{ABCD} \nb^{C} \cG^{rAB} \right)
\delta (rB^{rD}) \right\}  \ .
\eeq
Now, introducing the following set of variables
\beq
Q_1^{\ A} &=& rD^{rA}\ ,\\
Q_2^{\ A} &=& rB^{rA}\ ,\\
\Pi^1_{\ D} &=&  \frac{r}{2} \Delta_1^{-1}\epsilon_{ABCD} \nb^{C} B^{AB}\ ,\\
\Pi^2_{\ D} &=& -  \frac{r}{2} \Delta_1^{-1}\epsilon_{ABCD} \nb^{C} D^{AB}\ ,
\eeq
eq. (\ref{7-2}) simplifies to
\beq         \label{8-2}
- \delta \cH_{sym} &=&  \int_{V}\Lambda_2 \left\{ \left(
\dot{{\Pi}}^1_{\ A} \delta
Q_1^{\ A} -
\dot{Q}_1^{\ A} \delta \Pi^1_{\ A} \right)
- \left( \dot{{\Pi}}^2_{\ A} \delta
Q_2^{\ A} -
\dot{Q}_2^{\ A} \delta \Pi^2_{\  A} \right) \right\} \nonumber \\ &+&
 \int_{\p V}\Lambda_2 \left( \chi^1_{\ A} \delta Q_1^{\ A} + \chi^2_{\ A}
\delta Q_2^{\ A} \right) \ ,
\eeq
where we introduced the boundary momenta:
\beq
\chi^1_{\ A} &=& \frac{2}{r} A_{0A}\ ,\\
\chi^2_{\ D} &=& - \frac{r}{2} \Delta_1^{-1}\epsilon_{ABCD} \nb^{C} G^{rAB}\ .
\eeq
In (\ref{8-2}) we defined
\beq
Q_{l\, A} := g_{AB}\ Q_l^{\ B}\ , \ \ \ \
\Pi^{l\, A} := g^{AB}\ \Pi^l_{\ B}\ .
\eeq
Note the crucial difference between (\ref{8-2}) and (\ref{8-1}): the sign
``$+$'' in (\ref{8-1}) is replaced by ``$-$'' in (\ref{8-2}).

For a Maxwell theory one obtains
\beq
\cH_{sym}^{Maxwell} &=& \frac 14 \int_V \Lambda_2 \sum_{l=1}^2 \left\{
\frac{1}{r^2}  Q_l^{\ A} Q_{l\, A} -
\frac{1}{r^4} \p_r (r^3Q_{l\, A}) \Delta_1^{-1} \p_r(rQ_l^{\ A}) -
\Pi^{l\, A} \Delta_1 \Pi^l_{\ A} \right\}\
\eeq
and, therefore
\beq    \label{chi-12}
\chi^l_{\ A} = \frac{1}{r^3}\Delta_1^{-1} \p_r(r^3Q_{l\, A})\ , \ \ \ \ \
\ l=1,2\ .
\eeq

\subsection{Canonical symmetries}

The symplectic form $-\int \delta \cD^{ij} \wedge \delta A_{ij}$ rewritten
in terms of
$Q$'s and $\Pi$'s have the following form \cite{ROMP}:
\begin{equation}  \label{Om-2}
\Omega = \mbox{Im}\, \int \Lambda_2 \, \delta\Pi^A \wedge \delta {Q}_A\ ,
\end{equation}
where we introduced a complex notation
\beq
Q_A &=& Q^1_{\ A} +iQ^2_{\ A}\ , \\
\Pi^A &=& i(\Pi_1^{\ A} + i\Pi_2^{\ A})\ .
\eeq
The form (\ref{Om-2}) contrary to (\ref{Om-1}) is invariant only under
 the following  transformations:
\beq        \label{33}
Q_A &\rightarrow& \cosh\lambda\, Q_A + \sinh\lambda\, \overline{Q}_A\ ,
\eeq
and the same rule for $\Pi^A$. It is easy to see that these transformations
form the group $SO(1,1)$. In terms of $D^{ij}$ and $B^{ij}$, (\ref{33})
reads:
\beq                   \label{IV}
D^{ij} &\rightarrow & e^{\lambda}\,D^{ij} \ ,\nonumber\\
B^{ij} &\rightarrow & e^{-\lambda}\, B^{ij}\ .
\eeq
The canonical generator corresponding to (\ref{33}) has the
following form:
\beq   \label{G4-2}
G_4 &=& - \int \Lambda_2\, (Q^1_{\ A}\Pi_1^{\ A} + Q^2_{\ A}\Pi_2^{\ A}) =
\mbox{Im}\, \int \Lambda_2\,
(\Pi^A\overline{Q}_A)\ .
\eeq

\subsection{Summary}

Contrary to the $p=1$ case the reduced variables $(Q_l^{\ A},\Pi^l_{\ A})$
do not solve completely the Gauss constraints $\p_i D^{ij} = \p_i
B^{ij}=0$. They fulfill the
following additional conditions \cite{ROMP}:
\beq
\nb_A Q_l^{\ A} = \nb^A \Pi^l_{\ A} = 0\ ,\ \ \ \ l=1,2.
\eeq
In the geometric language it means that $\star Q_l$ and $\star \Pi^l$ are
closed 3-forms on $S^4(r)$ ($\star$ denotes the Hodge dual defined {\em via}
$\epsilon^{ABCD}$). They are gauge-invariant and contain the entire
information about 2-forms $D^{ij}$ and $B^{ij}$. The dynamics is generated
by the  {\it
quasi-local} functional of  $Q$'s and $\Pi$'s.

  The ``symmetric'' dynamics defined by (\ref{8-2})
corresponds to the Dirichlet boundary condition for positions $Q_l$
whereas the ``canonical'' dynamics corresponds to the Dirichlet conditions
for $\chi^1$ and $Q_2$. But Dirichlet condition for $\chi^1_{\ A}$ is
equivalent to the Neumann condition for $\p_r\cD^r_{\ A}$
\beq
\int_{\p V}\Lambda_2 Q_1^{\ A} \delta \chi^1_{\ A} = \int_{\p V} r
\Delta^{-1}_1 Q_1^{\ A} \delta (\p_r \cD^r_{\ A})\ .
\eeq

\section{Coupling to the charged matter}
\label{MATTER}
\setcounter{equation}{0}

In the present Section we study the coupling of $p$-form electrodynamics
to the charged matter. We present parallel discussion for $p=1$ and $p=2$.
The general case is presented in Appendix~C.

\subsection{$p=1$}

Consider a $1$-form electromagnetism interacting with the charged matter field
$\Phi$ (for simplicity let $\Phi$ be a complex scalar field). In the presence
of charged matter the Lagrangian generating formula (\ref{gen1}) has to be
replaced by:
\beq   \label{mat-1}
- \delta \cL =
 \partial_\nu (G^{\nu\mu}\delta
A_{\mu} +  \cP^\nu\delta\Phi)\ ,
\eeq
where the matter ``momentum''
\begin{equation}   \label{MM}
\cP^\nu = -\frac{\partial \cL}{\partial (\partial_\nu\Phi)}\ .
\end{equation}
Because $\cL$ should define a gauge-invariant theory let us assume that
there is a group of gauge transformations $U_\Lambda$ parameterized by a
a function ($0$-form) $\Lambda$
acting in the following
way:   $A_\mu\rightarrow A_\mu + \p_\mu\Lambda$ and $\Phi\rightarrow
U_\Lambda(\Phi)$.

Now, the target space of the matter field $\Phi$ may be reparameterized
$\Phi = (\vp,U)$ in such a way that, a parameter $U$ is gauge invariant
and $\vp$ is the phase undergoing the following gauge transformation: $\vp
\rightarrow \vp + \Lambda$. For the scalar (complex) field one has:
 $U := |\Phi|$ and the
 $\varphi = \mbox{Arg}\, \Phi$.  Therefore, the matter part in
(\ref{mat-1}) may be rewritten as follows:
\begin{equation}
\cP^\nu\delta \Phi =
J^{\nu}\delta\varphi +
p^\nu\delta U\ .
\end{equation}
Gauge invariance of the theory means that the gauge dependent quantities,
i.e. $A_\mu$ and $\vp$, enter into $\cL$ {\it via} the gauge-invariant
combinations only:
\beq
\cL = \cL(F_{\mu\nu},D_\mu\vp,U,\p_\mu U)\ ,
\eeq
where
\beq
D_\mu \vp := \p_\mu\vp - A_\mu\
\eeq
denotes a covariant derivative of $\vp$. This implies, that the
momentum $J^\mu$ canonically conjugated to $\vp$ is equal to the electric
current
\beq   \label{f-e1}
J^\mu = - \frac{\p \cL}{\p (\p_\mu\vp)} =  \frac{\p \cL}{\p A_\mu} = -
\p_\nu G^{\nu\mu}\ .
\eeq
Now, instead of (\ref{2-1}) one has
\beq  \label{2-1M}
- \delta\ \int_{V} \cL &=&   \int_{V} \partial_0\left( \cD^{i}
\delta A_{i} + \rho\delta\vp + p^0\delta U\right) \nonumber \\
&+& \int_{\partial V} \left(- \cD^{r}\delta
A_{0}  +  \cG^{rB}\delta A_{B} + J^r\delta\vp + p^r\delta U\right) \ ,
\eeq
with $\rho := J^0$.
Performing the set of Lagrange transformations between: 1) $\cD^k$ and
$\dot{A}_k$, 2) $\rho$ and $\dot{\vp}$, 3) $\pi := p^0$ and $\dot{U}$ in
the volume $V$, and between $\cD^r$ and $A_0$ at the boundary $\p V$, one
obtains the following generalization of (\ref{4-1}):
\beq   \label{4-1M}
- \delta {\cH}_{sym} &=& - \int_{V} \left\{ \left(
\dot{{\cD}}^{i}
\delta
A_{i} - \dot{A}_{i} \delta \cD^{i} \right) +
\left( \dot{\rho}\delta \vp - \dot{\vp} \delta \rho \right) +
\left( \dot{\pi}\delta U - \dot{U} \delta \pi \right) \right\} \nonumber\\
&-& \int_{\partial V} \left\{ A_{0}\delta \cD^{r} + \cG^{rB}\delta A_{B}
+ J^r\delta\vp + p^r\delta U \right\} \ ,
\eeq
where the ``symmetric'' Hamiltonian of the interacting electromagnetic
field and the charged matter represented by $\Phi$ reads:
\beq   \label{HM-1}
\cH_{sym} = \int_{V} \left( - \cD^{i}\dot{A}_{i} - \rho\dot{\vp} -
\pi\dot{U} - \cL + \p_k(A_0\cD^k) \right)\ .
\eeq
Now, using
\beq   \label{g1}
\p_k \cD^k = \rho\ ,
\eeq
implied by (\ref{f-e1}), one gets the following formula for $\cH_{sym}$:
\beq   \label{HM-1a}
\cH_{sym} = \int_{V} \left(  \cD^{i}E_{i} - \rho D_0{\vp} -\pi\dot{U} -
\cL \right)\ .
\eeq
Note, that the gauge-dependent phase $\vp$ enters into $\cH_{sym}$ {\it
via} the gauge-invariant combination $D_0 \vp$ only. Moreover, due to
(\ref{g1}), we may rewrite the dynamical part for $\vp$ in (\ref{4-1M}) as
follows:
\beq
\int_{V} \left( \dot{\rho}\delta\vp - \dot{\vp}\delta\rho \right) &=&
\int_{V} \left( - \dot{\cD}^k \delta (\p_k\vp) + (\p_k\dot{\vp})\delta
\cD^k \right) 
+ \int_{\p V} \left( \dot{\cD}^r \delta\vp - \dot{\vp}\delta\cD^r
\right)\ .
\eeq
Now, the $\dot{\cD}^r$ at the boundary may be easily eliminated by the
field equations (\ref{f-e1})
\beq     \label{f?}
\dot{\cD}^r = -\p_0 \cG^{r0} = \p_\mu \cG^{\mu r} - \p_A \cG^{Ar} = - J^r
+ \p_A \cG^{rA}\ .
\eeq
Introducing a hydrodynamical variables:
\beq
V_\mu := - D_\mu \vp\ ,
\eeq
we may rewrite finally (\ref{4-1M}) as follows:
\beq   \label{5-1M}
- \delta {\cH}_{sym} &=& - \int_{V} \left\{ \left(
\dot{{\cD}}^{i}
\delta
V_{i} - \dot{V}_{i} \delta \cD^{i} \right) +
\left( \dot{\pi}\delta U - \dot{U} \delta \pi \right) \right\} \nonumber\\
&-& \int_{\partial V} \left\{ V_{0}\delta \cD^{r} + \cG^{rB}\delta V_{B}
+ p^r\delta U \right\} \ ,
\eeq
i.e. (\ref{5-1M}) has exactly the same form as (\ref{4-1}) with $A_\mu$
replaced by the gauge-invariant $V_\mu$ and supplemented by the
gauge-invariant canonical pair $(U,\pi)$ together with the boundary
momentum $p^r$.

\subsection{$p=2$}

Now, consider a $2$-form electromagnetism interacting with the charged
matter field
$\Phi_\mu$ (for simplicity let $\Phi_\mu$ be a complex vector field). In
the presence
of charged matter the Lagrangian generating formula (\ref{gen-2}) has to be
replaced by:
\beq   \label{mat-2}
- \delta \cL =
 \partial_\nu (G^{\nu\mu\lambda}\delta
A_{\mu\lambda} +  \cP^{\nu\mu}\delta\Phi_\mu)\ ,
\eeq
where the matter ``momentum''
\begin{equation}
\cP^{\nu\mu} = -2 \frac{\partial \cL}{\partial (\partial_{[\nu}\Phi_{\mu]})}\ .
\end{equation}
Because $\cL$ should define a gauge-invariant theory let us assume that
there is a group of gauge transformations $U_\Lambda$ parameterized by a
a $1$-form $\Lambda$
acting in the following
way:   $A\rightarrow A + d\Lambda$ and $\Phi\rightarrow U_\Lambda(\Phi)$.

Now, the target space of the matter field $\Phi_\mu$ may be reparameterized
$\Phi_\mu = (\vp_\mu,U_\mu)$ in such a way that a 1-form $U_\mu$ is gauge
invariant and a 1-form $\vp_\mu$ is the phase undergoing the following
gauge transformation: $\vp
\rightarrow \vp + \Lambda$. For the vector (complex) field one has:
 $U_\mu := |\Phi_\mu|$ and
 $\varphi_\mu = \mbox{Arg}\, \Phi_\mu$.  Therefore, the matter part in
(\ref{mat-2}) may be rewritten as follows:
\begin{equation}
\cP^{\nu\mu}\delta \Phi_\mu =
J^{\nu\mu}\delta\varphi_\mu +
p^{\nu\mu}\delta U_\mu\ .
\end{equation}
Gauge invariance of the theory means that the gauge dependent quantities,
i.e. $A_{\mu\nu}$ and $\vp_\mu$, enter into $\cL$ {\it via} the gauge-invariant
combinations only:
\beq
\cL = \cL(F_{\mu\nu\lambda},D_\mu\vp_\nu,U_\mu,\p_\mu U_\nu)\ ,
\eeq
where
\beq
D_\mu \vp_\nu :=  \frac 12\, \p_{[\mu}\vp_{\nu]}  - A_{\mu\nu}
\eeq
denotes a ``covariant derivative'' of $\vp_\nu$. This implies, that the
momentum $J^{\mu\lambda}$ canonically conjugated to $\vp_\lambda$ is equal
to the electric current
\beq   \label{f-e2}
J^{\mu\lambda} = - 2\,\frac{\p \cL}{\p (\p_{[\mu}\vp_{\lambda]})} =  2\,\frac{\p
\cL}{\p A_{\mu\lambda}} = -
\p_\nu G^{\nu\mu\lambda}\ .
\eeq
Now, instead of (\ref{2-2}) one has
\beq  \label{2-2M}
- \delta\ \int_{V} \cL &=&   \int_{V} \partial_0\left( - \cD^{ij}
\delta A_{ij} - \rho^k\delta\vp_k  + \pi^{k}\delta U_k\right) \nonumber \\
&+& \int_{\partial V} \left( 2\cD^{rA}\delta
A_{0A}  +  \cG^{rAB}\delta A_{AB} + \rho^r\delta\vp_0 + J^{rA}\delta\vp_A
- \pi^{r}\delta U_0 + p^{rA}\delta U_A\right) \ ,
\eeq
with $\rho^k := J^{k0}$ (it defines a 1-form charge density on 5-dim.
hyperplane $\Sigma$) and $\pi^k := p^{0k}$.
Now, to pass to the Hamiltonian picture one has to
perform  the following Legendre transformations between: 1) $\cD$ and
$\dot{A}$, 2) $\rho$ and $\dot{\vp}$, 3) $\pi$ and $\dot{U}$ in
the volume $V$, and between 4) $\cD^r$ and $A_0$, 5) $\rho^r$ and $\vp_0$
and 6) $\pi^r$ and $U_0$ at the boundary $\p V$. One
obtains the following generalization of (\ref{4-2}):
\beq   \label{4-2M}
- \delta {\cH}_{sym} &=&  \int_{V} \left\{ \left(
\dot{{\cD}}^{ij}
\delta
A_{ij} - \dot{A}_{ij} \delta \cD^{ij} \right) +
\left( \dot{\rho}^k\delta \vp_k - \dot{\vp}_k \delta \rho^k \right) -
\left( \dot{\pi}^k\delta U_k - \dot{U}_k \delta \pi^k \right) \right\}
\nonumber\\
&-& \int_{\partial V} \left\{-2 A_{0A}\delta \cD^{rA} + \cG^{rAB}\delta A_{AB}
- \vp_0\delta\rho^r  + J^{rA}\delta\vp_A + U_0\delta \pi^r + p^{rA}\delta
U_A \right\} \ ,
\eeq
where the ``symmetric'' Hamiltonian of the interacting electromagnetic
field and the charged matter represented by $\Phi_\mu$ reads:
\beq   \label{HM-2}
\cH_{sym} = \int_{V} \left\{ \cD^{ij}\dot{A}_{ij} + \rho^k\left(
\dot{\vp}_k - \p_k \vp_0 \right) - \pi^k\dot{U}_k -
 \p_k\left(2A_{0i}\cD^{ki} - U_0\pi^k\right) - \cL \right\}\ ,
\eeq
where we have used $\p_k\rho^k =0$.
Now, using
\beq   \label{g2}
\p_i \cD^{ik} = \rho^k\ ,
\eeq
one gets the following formula for $\cH_{sym}$:
\beq   \label{HM-2a}
\cH_{sym} = \int_{V} \left( \frac 12 \cD^{ij}E_{ij} + 2\rho^k D_0{\vp}_k -
\pi^k\dot{U}_k - \cL  + \p_k (\pi^k U_0)\right)\ .
\eeq
Note, that the gauge-dependent phase $\vp_\mu$ enters into $\cH_{sym}$ {\it
via} the gauge-invariant combination $D_0 \vp_\mu$. Moreover, due to
(\ref{g2}), we may rewrite the dynamical part for $\vp_\mu$ in (\ref{4-2M}) as
follows:
\beq
\int_{V} \left( \dot{\rho}^k\delta\vp_k - \dot{\vp}_k\delta\rho^k \right) &=&
\int_{V} \left( - \dot{\cD}^{ik} \delta (\p_i\vp_k) +
(\p_i\dot{\vp}_k)\delta \cD^{ik} \right) 
+ \int_{\p V} \left( \dot{\cD}^{rA} \delta\vp_A - \dot{\vp}_A\delta\cD^{rA}
\right)\ .
\eeq
Now, the term  $\dot{\cD}^{rA}$ at the boundary may be easily eliminated
by the field equations (\ref{f-e2})
\beq     \label{f??}
\dot{\cD}^{rA} =  J^{rA} + \p_B \cG^{rAB}\ .
\eeq
Introducing  hydrodynamical variables:
\beq
V_{\mu\nu} := - D_\mu \vp_\nu\ ,
\eeq
we may rewrite finally (\ref{4-2M}) as follows:
\beq   \label{5-2M}
- \delta {\cH}_{sym} &=&  \int_{V} \left\{ \left(
\dot{{\cD}}^{ij}
\delta
V_{ij} - \dot{V}_{ij} \delta \cD^{ij} \right) -
\left( \dot{\pi}^k\delta U_k - \dot{U}_k \delta \pi^k \right) \right\}
\nonumber\\
&-& \int_{\partial V} \left\{ -2 V_{0A}\delta \cD^{rA} + \cG^{rAB}\delta
V_{AB} -U_0\delta\pi^r + p^{rA}\delta U_A \right\} \ ,
\eeq
i.e. (\ref{5-1M}) has exactly the same form as (\ref{4-2}) with $A_{\mu\nu}$
replaced by the gauge-invariant 2-form $V_{\mu\nu}$ and supplemented by the
gauge-invariant canonical pair $(U_k,\pi^k)$ together with the boundary
momenta $U_0$ and $p^{rA}$. All gauge-dependent terms dropped out.

\section*{Appendices}

\appendix

\def\theequation{\thesection.\arabic{equation}}

\section{Notation}
\setcounter{equation}{0}

Consider a $p$-form potential $A$ defined in the $D=2p+2$ dimensional Minkowski
space-time ${\cal M}^{2p+2}$ with the signature of the metric tensor
$(-,+,...,+)$.
The corresponding field tensor is defined as a $(p+1)$-form by $F=dA$:
\beq   \label{F}
F_{\mu_1...\mu_{p+1}} = \partial_{[\mu_1}A_{\mu_2...\mu_{p+1}]}\ ,
\eeq
where the antisymmetrization is defined by $X_{[kl]} :=
X_{kl} - X_{lk}$. Having a Lagrangian
$\cL$ of the theory one defines another $(p+1)$-form $G$ as follows:
\beq   \label{const}
\cG^{\mu_1...\mu_{p+1}} = - (p+1)!\frac{\partial \cL}{\partial
F_{\mu_1...\mu_{p+1}}}\ .
\eeq
Now one may define the electric and magnetic intensities and inductions in the
obvious way:
\beq   \label{E}
E_{i_1...i_p} &=& F_{i_1...i_p0}\ ,\\   \label{B}
B_{i_1...i_p} &=& \frac{1}{(p+1)!}\
\epsilon_{i_1...i_pj_1...j_{p+1}} F^{j_1...j_{p+1}}\ , \\   \label{D}
 \cD_{i_1...i_p} &=& \cG_{i_1...i_p0}\ ,\\         \label{H}
\cH_{i_1...i_p} &=& \frac{1}{(p+1)!} \
\epsilon_{i_1...i_pj_1...j_{p+1}} \cG^{j_1...j_{p+1}}\ ,
\eeq
where the indices $i_1,i_2,...,j_1,j_2,...$ run from 1 up to $2p+1$ and
$\epsilon_{i_1i_2...i_{2p+1}}$ is the L\'evi-Civita tensor in $2p+1$
dimensional Euclidean space, i.e. a space-like hyperplane $\Sigma$ in the
Minkowski space-time.
The field equations are given by the Bianchi identities $dF=0$, or
in components
\beq   \label{Bianchi}
\partial_{[\lambda}F_{\mu_1...\mu_{p+1}]} =0\ ,
\eeq
and the {\em true} dynamical equations $d\star \cG=0$, or equivalently
\beq  \label{dynamical}
\partial_{[\lambda}\star \cG_{\mu_1...\mu_{p+1}]} =0\ ,
\eeq
 where the Hodge star
operation in ${\cal M}^{2p+2}$ is defined by:
\beq
\star X^{\mu_1...\mu_{p+1}} = \frac{1}{(p+1)!} \ \eta^{\mu_1...\mu_{p+1}
 \nu_1...\nu_{p+1}}\ X_{\nu_1...\nu_{p+1}}\
\eeq
and $\eta^{\mu_1\mu_2...\mu_{2p+2}}$ is the covariantly constant volume form
in the Minkowski space-time. Note, that $\epsilon^{i_1...i_{2p+1}} :=
\eta^{0i_1...i_{2p+1}}$.
In terms of electric and magnetic fields defined
in (\ref{E})--(\ref{H}) the field equations (\ref{Bianchi})--(\ref{dynamical})
have the following form:
\beq               \label{dB}
\partial_0 B^{i_1...i_p} &=& (-1)^p \frac{1}{p!}\
\epsilon^{i_1...i_pkj_1...j_p}\
\nb_k E_{j_1...j_p}\ ,\\    \label{B1}
\nb_{i_1} B^{i_1...i_p} &=&0\ ,\\        \label{dD}
\partial_0 \cD^{i_1...i_p} &=&  \frac{1}{p!}\
\epsilon^{i_1...i_pkj_1...j_p}\
\nb_k \cH_{j_1...j_p}\ ,\\              \label{D1}
\nb_{i_1} \cD^{i_1...i_p} &=&0\ ,
\eeq
where $\nb_k$ denotes the covariant derivative on $\Sigma$ compatible with the
metric $g_{kl}$ induced from ${\cal M}^{2p+2}$. The
L\'evi-Civita tensor density satisfies
$\epsilon_{12...2p+1} = \sqrt{g}$,
with $g=\det(g_{kl})$.

\section{General $p$-form theory without matter}   \label{GENERAL}
\setcounter{equation}{0}

\subsection{Generating formula}

For an arbitrary $p$ the formulae (\ref{gen1}) and (\ref{gen-2}) generalize
to:
\beq   \label{gen}
- \delta \cL =  (\partial_\nu \cG^{\nu\mu_1...\mu_p}\delta
A_{\mu_1...\mu_p}) =  (\partial_\nu \cG^{\nu\mu_1...\mu_p})\delta
A_{\mu_1...\mu_p} +  \cG^{\nu\mu_1...\mu_p} \delta (\partial_\nu
A_{\mu_1...\mu_p}) \ .
\eeq
The formula (\ref{gen}) implies the following definition of ``momenta'':
\beq   \label{cG-p}
\cG^{\mu_1...\mu_{p+1}} =
 - (p+1)!\,\frac{\partial \cL}{\partial
F_{\mu_1...\mu_{p+1}}}\ .
\eeq
Moreover,  (\ref{gen}) generates  dynamical (in general nonlinear) field
equations
\beq   \label{f-eqs-p}
\partial_\nu \cG^{\nu\mu_1...\mu_p} = - \cJ^{\mu_1...\mu_p}\ ,
\eeq
where the external $p$-form current reads:
\beq            \label{J-p}
 \cJ^{\mu_1...\mu_p} =
 p!\, \frac{\partial \cL}{\partial A_{\mu_1...\mu_p}}\ .
\eeq
Let us start with  $\cJ=0$ and discuss a general $p$-form
charged matter in Appendix~C.
 To obtain the Hamiltonian description of the field dynamics
let us integrate equation (\ref{gen}) over a (2$p$+1)-dimensional volume $V$
contained in the constant-time hyperplane $\Sigma$:
\beq   \label{1}
- \delta\, \int_{V} \cL = \int_{V} \partial_0( \cG^{0i_1...i_p}\delta
A_{i_1...i_p}) + \int_{\partial V} \cG^{\perp\mu_1...\mu_p}\delta
A_{\mu_1...\mu_p}\ ,
\eeq
where $\perp$ denotes the component orthogonal to the $2p$-dimensional
boundary $\p V$.
 To simplify our
notation let us introduce the spherical coordinates on $\Sigma$:
\beq   \label{sp}
x^{2p+1} &=& r\ , \ \ \ \ x^A = \vp_A\ ;\ \ \ \
A=1,2,...,2p\ , \eeq
where $\vp_1,\vp_2,...,\vp_{2p}$ denote
spherical angles (to enumerate angles we shall use capital letters
$A,B,C,...$). The metric tensor $g_{ij}$ is diagonal and has the following
form:
\beq   \nonumber
g_{11} &=& r^2 \sin^2\vp_2\sin^2\vp_3...\sin^2\vp_{2p}\ ,\\ \nonumber
g_{22} &=& r^2 \sin^2\vp_3\sin^2\vp_4...\sin^2\vp_{2p}\ ,\\
       &\vdots&  \\  \nonumber
g_{2p-1,2p-1} &=& r^2\sin^2\vp_{2p-1}\sin^2\vp_{2p}\, \\ \nonumber
g_{2p,2p} &=& r^2\sin^2\sin_{2p}\ , \\ \nonumber
g_{rr} &=& r^2\ .
\eeq
Therefore, the volume form
\beq
\Lambda_p = \sqrt{\det(g_{ij})} = r^{2p}
\sin\vp_2\sin^2\vp_3...\sin^{2p-2}\vp_{2p-1}\sin^{2p-1}\vp_{2p}\ .
\eeq
Let $V$ be
 a (2$p$+1)--dim. ball with a finite radius $R$. In such a coordinate system
the formula
(\ref{1}) takes the following form:
\beq  \label{2}
\delta\ \int_{V} \cL &=& (-1)^{p} \int_{V} \partial_0( \cD^{i_1...i_p}
\delta A_{i_1...i_p}) - (-1)^p \int_{\partial V} p\,\cD^{rA_2...A_p}\delta
A_{0A_2...A_p}          \nonumber \\ &-&
\int_{\partial V} \cG^{rB_1...B_p}\delta A_{B_1...B_p}\ ,
\eeq
where
\begin{equation}
\cD_{i_1...i_p} = \cG_{i_1...i_p0}
\end{equation}
denotes the $p$-form electric induction density.
Now, performing the Legendre transformation between induction $p$-form
$\cD^{i_1...i_p}$ and $\dot{A}_{i_1...i_p}$ one obtains the following
Hamiltonian formula:
\beq   \label{3}
- \delta \cH_{can} &=& (-1)^{p} \int_{V} \left( \dot{{\cD}}^{i_1...i_p} \delta
A_{i_1...i_p} - \dot{A}_{i_1...i_p} \delta \cD^{i_1...i_p} \right)
- (-1)^p \int_{\partial V} p\, \cD^{rA_2...A_p}\delta A_{0A_2...Ap}
 \nonumber \\
&-& \int_{\partial V} \cG^{rB_1...B_p}\delta A_{B_1...B_p}\ ,
\eeq
where the canonical Hamiltonian
\beq   \label{Hp}
\cH_{can} = \int_{V} \left( (-1)^{p} \cD^{i_1...i_p}\dot{A}_{i_1...i_p} - \cL
\right)\ .
\eeq
Equation (\ref{3}) generates an infinite-dimensional Hamiltonian system in
the phase space ${\cal P}_p = (\cD^{i_1...i_p},A_{i_1...i_p})$
fulfilling Dirichlet boundary conditions for the $p$-form potential
$A_{i_1...i_p}$: $A_{0A_2...A_p}|\partial V$ and $A_{A_1A_2...A_p}|\partial
V$. From the mathematical point of view this is the missing part of the
definition of the functional space. The Hamiltonian structure of a general
nonlinear $p$-form electrodynamics described above is mathematically well
defined, i.e. a mixed Cauchy problem (Cauchy data given on $\Sigma$ and
Dirichlet data given on $\p V \times \bR$) has a unique solution (modulo gauge
transformations which reduce to the identity on $\p V \times \bR$).

The presence of a $p$-dependent sign $(-1)^p$
follows from the $p$-dependence of the corresponding symplectic form:
\begin{equation}   \label{Omegap}
\Omega_p = \int_{V} \delta \cG^{0i_1...i_p} \wedge \delta A_{i_1...i_p} =
(-1)^{p+1} \int_{V}\delta\cD^{i_1...i_p}\wedge \delta{A}_{i_1...i_p}\ .
\end{equation}

There is, however, another way to describe the Hamiltonian evolution of
fields in the region $V$. Let us perform the Legendre transformation
between $\cD^{rA_2...A_p}$ and $A_{0A_2...A_p}$ at the boundary $\partial
V$. One obtains:
\beq   \label{4}
- \delta {{\cH}}_{sym} &=& (-1)^{p} \int_{V} \left(
\dot{{\cD}}^{i_1...i_p}
\delta
A_{i_1...i_p} - \dot{A}_{i_1...i_p} \delta \cD^{i_1...i_p} \right)
+ (-1)^p \int_{\partial V} pA_{0A_2...A_p}\delta \cD^{rA_2...A_p} \nonumber\\
&-& \int_{\partial V} \cG^{rB_1...B_p}\delta A_{B_1...B_p}\ ,
\eeq
where the new ``symmetric'' Hamiltonian
\beq   \label{Hpo}
{{\cH}}_{sym} = \cH_{can}  - (-1)^p \int_{\partial V}p\, \cD^{rA_2...A_p}
{A}_{0A_2...A_p} \ .
\eeq
Observe, that formula (\ref{4}) defines the Hamiltonian evolution but on a
different phase space. In order to kill boundary terms in (\ref{4}) one has to
control on $\partial V$: $\cD^{rA_2...A_p}$ (instead of $A_{0A_2...A_p}$) and
$A_{B_1...B_p}$. We stress that from the mathematical point of view both
descriptions are equally good and an additional physical argument has to be
given if we want to choose one of them as more fundamental.

\subsection{Canonical vs. symmetric energy}

Now, let us discuss the relation between $\cH_{can}$ and ${{\cH}}_{sym}$
defined by (\ref{Hp}) and (\ref{Hpo}) respectively. One has:
\beq   \label{Hpo1}
{{\cH}}_{sym} &=& \cH_{can}  - (-1)^p \int_{\partial V} p\,\cD^{rA_2...A_p}
{A}_{0A_2...A_p}
= \cH - (-1)^p \int_{V} p\,\partial_k\left( \cD^{ki_2...i_p}
{A}_{0i_2...i_p} \right) \nonumber\\
&=& \int_{V} \left\{ (-1)^p \cD^{i_1...i_p}\dot{A}_{i_1...i_p} - \cL +
(-1)^p p\, \left(
{A}_{0i_2...i_p}\partial_k\cD^{ki_2...i_p} + \cD^{ki_2...i_p}
\partial_k{A}_{0i_2...i_p} \right) \right\}    \nonumber\\
&=& \int_{V} \left( \frac{1}{p!}\,\cD^{i_1...i_p}E_{i_1...i_p} -
\cL\right)\ ,
\eeq
where the  $p$-form electric field is defined by
\begin{equation}
E_{i_1...i_p} = F_{i_1...i_p0} = \p_{[i_1}A_{i_2...i_p0]}\ .
\end{equation}
Therefore,
 $\cH_{sym} = \int_{V} T^{00}_{sym}$ and ${\cH}_p =
\int_{V} T^{00}_{can}$, where
\beq  \label{Ts}
T^{\mu\nu}_{sym} &=& \frac{1}{p!} F^{\mu\nu_1...\nu_p} G^\nu_{\ \nu_1...\nu_p}
+ g^{\mu\nu}\cL\ ,  \\
T^{\mu\nu}_{can} &=& \p^\mu A^{\nu_1...\nu_p} G^{\nu}_{\ \nu_1...\nu_p}
+ g^{\mu\nu}\cL\ .
\eeq
Obviously, for the Maxwell theory one has:
\beq
\cH^{Maxwell}_{sym} = \frac{1}{2p!} \int \left( \cD^{i_1...i_p}D_{i_1...i_p}
+ \cB^{i_1...i_p}B_{i_1...i_p}  \right)\ .
\eeq

\subsection{Reduction of the generating formula}

Any geometrical object on $(2p+1)$--dimensional hyperplane $\Sigma$ may be
decomposed into the radial and tangential components, e.g. a $p$-form gauge
potential $A_{i_1...i_p}$ decomposes into the radial $A_{rA_2...A_p}$ and
tangential $A_{A_1...A_p}$. On each sphere $2p$-dimensional sphere
$S^{2p}(r)$, $A_{rA_2...A_p}$ defines a $(p-1)$--form whereas $A_{A_1...A_p}$
a $p$-form. Now, any $p$-form on $S^{2p}(r)$ may be further decomposed into
``longitudinal'' and ``transversal'' parts:
\begin{equation}   \label{dec}
A_{A_1...A_p} = \nabla_{[A_1}u_{A_2...A_p]} +
\epsilon_{A_1...A_pB_1...B_p}\nabla^{B_1} v^{B_2...B_p}\ ,
\end{equation}
where $\epsilon_{A_1...A_pB_1...B_p}$ denotes the L\'evi-Civita tensor density
on $S^{2p}(r)$ such that $\epsilon_{12...2p}= \Lambda_p$.
Both $u$ and $v$ are $(p-1)$--forms on $S^{2p}(r)$. Now, using (\ref{dec}) and
integrating by parts one gets:
\beq  \label{4a}
\lefteqn{\int_{V} \left(
\dot{{\cD}}^{i_1...i_p}
\delta
A_{i_1...i_p} - \dot{A}_{i_1...i_p} \delta \cD^{i_1...i_p}
\right)        }
\nonumber\\    &=&
\int_{V} p\,  \left\{ \left(
\dot{{\cD}}^{rA_2...A_p}
\delta
A_{rA_2...A_p} - \dot{A}_{rA_2...A_p} \delta \cD^{rA_2...A_p} \right)
\right. \nonumber\\    &+&
 p!\, \left[
\left(\p_r\dot{{\cD}}^{rA_2...A_p}\right)
\delta
u_{A_2...A_p} - \dot{u}_{A_2...A_p} \delta \left(\p_r\cD^{rA_2...A_p}\right)
\right]
\nonumber\\    &-& \left.
\epsilon_{A_1...A_pB_1...B_p} \left[
\left(\nb^{B_1}\dot{{\cD}}^{A_1...A_p}\right)
\delta
v^{B_2...B_p} - \dot{v}^{B_2...B_p} \delta
\left(\nb^{B_1}\cD^{A_1...A_p}\right)
\right] \right\}
\ ,
\eeq
where we have used the Gauss law
\begin{equation}
\nb_{A_1} \cD^{A_1...A_p} = - \p_r \cD^{rA_2...A_p}\ .
\end{equation}
Moreover, due to (\ref{dec})
\beq    \label{5}
\lefteqn{\int_{\p V} {\cG}^{rA_1...A_p} \delta A_{A_1...A_p} }
\nonumber \\   &=&
 \int_{\p V} \left\{ (-1)^p p!\, \dot{\cD}^{rA_2...A_p}
\delta u_{A_2...A_p}
-   \left(\epsilon_{A_1...A_pB_1...B_p} \nb^{B_1}
\cG^{rA_1...A_p}\right)\delta v^{B_2...B_p}\right\}\ .
\eeq
In deriving (\ref{5}) we have used
\begin{equation}
\nb_{A_1} \cG^{A_1...A_pr} = - \dot{{\cD}}^{rA_2...A_p}\ ,
\end{equation}
which follows from the field equations $\nb_{A_1} \cG^{A_1...A_pr} + \p_0
\cG^{0A_2...A_pr} = 0$.
Now, taking into account (\ref{4a}) and (\ref{5}) the generating formula
(\ref{4}) may be rewritten in the following way:
\beq           \label{6}
- \delta \cH_{sym} &=& (-1)^p\int_{V} \left\{ \left[
\dot{{\cD}}^{rA_2...A_p} \delta ( pA_{rA_2...A_p} - p!\p_r u_{A_2...A_p} )
 \right. \right. \nonumber \\  &-&  \left.
( p\dot{A}_{rA_2...A_p} - p!\p_r \dot{u}_{A_2...A_p} ) \delta
\cD^{rA_2...A_p} \right] - \left[ (\epsilon_{A_1...A_pB_1...B_p} \nb^{B_1}
\dot{{\cD}}^{A_1...A_p}) \delta v^{B_2...B_p} \right. \nonumber \\
&-& \left. \left. \dot{v}^{B_2...B_p} \delta(\epsilon_{A_1...A_pB_1...B_p}
\nb^{B_1} \cD^{A_1...A_p} \right]  \right\} \nonumber \\
&+& (-1)^p \int_{\p V} (p\, A_{0A_2...A_p} - p!\, \dot{u}_{A_2...A_p}) \delta
\cD^{rA_2...A_p} \nonumber  \\
&+& \int_{\p V} (\epsilon_{A_1...A_pB_1...B_p} \nb^{B_1} \cG^{rA_1...A_p} )
\delta v^{B_2...B_p} \ .
\eeq
Note, that although $A_{rA_2...A_p}$, $A_{0A_2...A_p}$ and $u_{A_2...A_p}$ are
manifestly gauge-dependent, the combinations
$p\, A_{rA_2...A_p} - p!\, \p_r u_{A_2...A_p}$ and
$p\, A_{0A_2...A_p} - p!\, \p_0 {u}_{A_2...A_p}$
are gauge-invariant. To simplify our consideration we choose the special gauge
$u\equiv 0$, i.e. a $p$-form $A_{A_1...A_p}$ on $S^{2p}(r)$ is purely
transversal. This condition, due to (\ref{dec}), may be equivalently rewritten
as
\begin{equation} \label{gauge1p}
\nb_{A_1} A^{A_1...A_p} =0\ .
\end{equation}
Let us choose the same condition for the radial part
\begin{equation}    \label{gauge2p}
\nb_{A_2} A^{rA_2...A_p} =0\ .
\end{equation}
Assuming (\ref{gauge1p}) and (\ref{gauge2p}) one may show \cite{ROMP}
\beq  \label{Ar}
\Delta_{p-1} A^{rB_2...B_p} = (-1)^{p+1} \frac{r^2}{p\,p!}\,
\epsilon^{A_1...A_pB_1...B_p} \nb_{B_1} B_{A_1...A_p}\ ,
\eeq
where
\beq
\Delta_{p-1} = (p-1)! \left[ r^2 \nb_A\nb^A - (p^2-1)\right]
\eeq
equals to the Laplace-Beltrami operator on co-exact $(p-1)$-forms on
$S^{2p}(1)$ \cite{ROMP}. In the same way
\begin{equation}
 B^{rA_2...A_p} = - \frac{p!}{r^2}\, \Delta_{p-1} v^{A_2...A_p}\ .
\end{equation}
Finally, introducing
\beq
Q_1^{\ A_2...A_p} &=& D^{rA_2...A_p}\ ,\\
Q_2^{\ A_2...A_p} &=& B^{rA_2...A_p}\ ,\\    \label{pi1B}
\Pi^1_{\ B_2...B_p} &=& \frac{r}{p!}\, \Delta^{-1}_{p-1} \left(
\epsilon_{A_1...A_pB_1...B_p} \nb^{B_1} B^{A_1...A_p}\right)\ ,\\
\label{pi2B}
\Pi^2_{\ B_2...B_p} &=& - \frac{r}{p!}\, \Delta^{-1}_{p-1} \left(
\epsilon_{A_1...A_pB_1...B_p} \nb^{B_1} D^{A_1...A_p}\right) \ ,
\eeq
the formula (\ref{6}) simplifies to
\beq         \label{8}
- \delta \cH_{sym} &=&  \int_{V}\Lambda_p \left\{ \left(
\dot{{\Pi}}^1_{\ A_2...A_p}
\delta
Q_1^{\ A_2...A_p} -
\dot{Q}_1^{\ A_2...A_p}  \delta \Pi^1_{\ A_2...A_p}  \right) \right.
\nonumber\\ &+& \left.
 (-1)^{p+1} \left( \dot{{\Pi}}^2_{\ A_2...A_p}
\delta
Q_2^{\ A_2...A_p}  -
\dot{Q}_2^{\ A_2...A_p}   \delta \Pi^2_{\ A_2...A_p}  \right) \right\} \nonumber \\
&+&
 \int_{\p V}\Lambda_p \left( \chi^1_{\ A_2...A_p}  \delta Q_1^{\ A_2...A_p}
 + \chi^1_{\ A_2...A_p}
\delta Q_1^{\ A_2...A_p}
\right) \ ,
\eeq
where we introduced the boundary momenta:
\beq
\chi^1_{A_2...A_p} &=& (-1)^p \frac{p}{r}\, A_{0A_2...A_p}\ ,\\
\chi^2_{B_2...B_p}  &=& - \frac{r}{p!}\,
\Delta_1^{-1}\epsilon_{A_1...A_pB_1...B_p}
\nb^{B_1} G^{rA_1...A_p}\ .
\eeq
In the formula (\ref{8}) we have introduced:
\beq
Q_{l\, A_2...A_p} &:=& g_{A_2B_2}...g_{A_pB_p}\, Q_l^{\ B_2...B_p}\ ,\\
\Pi^{l\, A_2...A_p} &:=& g^{A_2B_2}...g^{A_pB_p}\, \Pi^l_{\ B_2...B_p}\ ,
\eeq
for $l=1,2$. For the Maxwell theory
\beq
\cH_{sym}^{Maxwell} &=& \frac{1}{2(p-1)!} \int_V \Lambda_p \sum_{l=1}^2
\left\{
\frac{1}{r^2}  Q_l^{\ A_2...A_p} Q_{l\, A_2...A_p}
- \Pi^{l\, A_2...A_p} \Delta_{p-1} \Pi^l_{\ A_2...A_p}
\right. \nonumber\\ &-&
\left.
\frac{1}{r^{2p}} \p_r (r^{2p-1}Q_{l\, A_2...A_p}) \Delta_{p-1}^{-1}
\p_r(rQ_l^{\ A_2...A_p})
\right\}\ ,
\eeq
and, therefore, the boundary momenta read:
\beq    \label{chi-12-p}
\chi^l_{\ A_2...A_p} = \frac{1}{r^{2p-1}}\Delta_1^{p-1} \p_r(r^{2p-1}Q_{l\,
A_2...A_p})\ , \ \ \ \ \ l=1,2.
\eeq

\subsection{Summary}

The {\it quasi-local} reduced variables $(Q_l^{\ A_2...A_p},\Pi^l_{\
A_2...A_p})$
fulfill the
following  conditions \cite{ROMP}:
\beq
\nb_{A_2} Q_l^{\ A_2...A_p} = \nb^{A_2} \Pi^l_{\ A_2...A_p} = 0\ ,\ \ \ \
l=1,2,
\eeq
which follow from the Gauss laws.
In the geometric language it means that $\star Q_l$ and $\star \Pi^l$ are
closed $(p+1)$--forms on $S^{2p}(r)$ ($\star$ denotes the Hodge dual defined
{\em via}
$\epsilon^{A_1...A_pB_1...B_p}$). They are gauge-invariant and contain the
entire
information about $p$-forms $D$ and $B$.

  The ``symmetric'' dynamics defined by (\ref{8})
corresponds to the Dirichlet boundary condition for positions $Q_l$
whereas the ``canonical'' dynamics corresponds to the Dirichlet conditions
for $\chi^1_{\ A_2...A_P}$ and $Q_2$.
But Dirichlet condition for $\chi^1_{\ A}$
is equivalent to the Neumann condition for $\p_r\cD^r_{\ A_2...A_p}$
\beq
\int_{\p V} \Lambda_p Q_1^{\ A_2...A_p} \delta \chi^1_{\ A_2...A_p}
= \int_{\p V} r \Delta^{-1}_{p-1}
Q_1^{\ A_2...A_p} \delta (\p_r \cD^r_{\ A_2...A_p})\ .
\eeq

\section{General $p$-form theory with matter}   \label{p-MATTER}
\setcounter{equation}{0}

Now, consider a $p$-form electromagnetism interacting with the charged
matter field
$\Phi$ (for simplicity let $\Phi$ be a complex $(p-1)$--form). In
the presence
of charged matter the Lagrangian generating formula (\ref{gen}) has to be
replaced by:
\beq   \label{mat}
- \delta \cL =
 \partial_\nu (G^{\nu\mu_1...\mu_p}\delta
A_{\mu_1\mu_p} +  \cP^{\nu\mu_2...\mu_p}\delta\Phi_{\mu_2...\mu_p})\ ,
\eeq
where the matter ``momentum''
\begin{equation}
\cP^{\mu_1\mu_2...\mu_p} = - p! \frac{\partial \cL}{\partial
(\partial_{[\mu_1}\Phi_{\mu_2...\mu_p]})}\ .
\end{equation}
Because $\cL$ should define a gauge-invariant theory let us assume that
there is a group of gauge transformations $U_\Lambda$ parameterized by a
a $p$-form $\Lambda$
acting in the following
way:   $A\rightarrow A + d\Lambda$ and $\Phi\rightarrow U_\Lambda(\Phi)$.

Now, the target space of the matter field $\Phi$ may be reparameterized
$\Phi = (\vp,U)$ in such a way that a $(p-1)$--form $U$ is gauge
invariant and a $(p-1)$--form $\vp$ is the phase undergoing the following
gauge transformation: $\vp
\rightarrow \vp + \Lambda$. For the  (complex) $(p-1)$--form one has:
 $U_{\mu_1...\mu_{p-1}} := |\Phi_{\mu_1...\mu_{p-1}}|$ and
 $\varphi_{\mu_1...\mu_{p-1}} = \mbox{Arg}\, \Phi_{\mu_1...\mu_{p-1}}$.
Therefore, the matter part in
(\ref{mat}) may be rewritten as follows:
\begin{equation}
\cP^{\mu_1...\mu_p}\delta \Phi_{\mu_2...\mu_p} =
J^{\mu_1...\mu_p}\delta\varphi_{\mu_2...\mu_p} +
p^{\mu_1...\mu_p}\delta U_{\mu_2...\mu_p}\ .
\end{equation}
Gauge invariance of the theory means that the gauge dependent quantities,
i.e. $A$ and $\vp$, enter into $\cL$ {\it via} the gauge-invariant
combinations only:
\beq
\cL =
\cL(F_{\mu_1...\mu_{p+1}},D_\nu\vp_{\mu_1...\mu_{p-1}},U_{\mu_1...\mu_{p-1}},\p_\nu
U_{\mu_1...\mu_{p-1}})\ ,
\eeq
where
\beq
D_\nu \vp_{\mu_1...\mu_{p-1}} :=  \frac 1p\,
\p_{[\nu}\vp_{\mu_1...\mu_{p-1}]}  - A_{\nu\mu_1...\mu_{p-1}}
\eeq
denotes a covariant derivative of $\vp_{\mu_1...\mu_{p-1}}$. This
implies, that the
momentum $J^{\mu_1...\mu_p}$ canonically conjugated to
$\vp_{\mu_1...\mu_{p}}$ is equal
to the electric current
\beq   \label{f-e}
J^{\mu_1...\mu_p} = - p!\,\frac{\p \cL}{\p
(\p_{[\mu_1}\vp_{\mu_2...\mu_p]})} =  p!\,\frac{\p
\cL}{\p A_{\mu_1...\mu_p}} = -
\p_\nu G^{\nu\mu_1...\mu_p}\ .
\eeq
Now, instead of (\ref{2}) one has
\beq  \label{2M}
\lefteqn{
\delta\ \int_{V} \cL =   \int_{V} \partial_0\left\{ (-1)^{p}
\cD^{i_1...i_p}\delta A_{i_1...i_p} + (-1)^{p}
\rho^{i_1...i_{p-1}}\delta\vp_{i_1...i_{p-1}}  - \pi^{i_1...i_{p-1}}\delta
U_{i_1...i_{p-1}} \right\} }
\nonumber \\
&-& \int_{\partial V} \left\{ (-1)^p p\, \cD^{rA_2...A_p}\delta
A_{0A_2...A_p}  +  \cG^{rA_1...A_p}\delta A_{A_1...A_p} +
(-1)^p (p-1) \rho^{rA_3...A_p}\delta\vp_{0A_3...A_p}  \right. \nonumber\\ &+&
\left. J^{rA_2...A_p}\delta\vp_{A_2...A_p}
- (p-1)\pi^{rA_3...A_p}\delta U_{0A_3...A_p} + p^{rA_2...A_p}\delta
U_{A_2...A_p}\right\} \ ,
\eeq
with $\rho^{i_1...i_{p-1}} := J^{i_1...i_{p-1}0}$ (it defines a
$(p-1)$--form charge density on $(2p+1)$--dim.
hyperplane $\Sigma$) and $\pi^{i_1...i_{p-1}}:= p^{0i_1...i_{p-1}}$.
Now, to pass to the Hamiltonian picture one has to
perform  the following Legendre transformations between: 1) $\cD$ and
$\dot{A}$, 2) $\rho$ and $\dot{\vp}$, 3) $\pi$ and $\dot{U}$ in
the volume $V$, and between 4) $\cD^r$ and $A_0$, 5) $\rho^r$ and $\vp_0$
and 6) $\pi^r$ and $U_0$ at the boundary $\p V$. One
obtains the following generalization of (\ref{4}):
\beq   \label{4M}
\lefteqn{
- \delta {\cH}_{sym} =  \int_{V} \left\{
(-1)^{p} \left( \dot{\cD}^{i_1...i_p}\delta A_{i_1...i_p}
- \dot{A}_{i_1...i_p} \delta \cD^{i_1...i_p} \right)
\right. } \nonumber\\ &+&
(-1)^{p}\left( \dot{\rho}^{i_1...i_{p-1}}\delta \vp_{i_1...i_{p-1}} -
\dot{\vp}_{i_1...i_{p-1}} \delta \rho^{i_1...i_{p-1}} \right)
  \nonumber\\  &-& \left.
\left( \dot{\pi}^{i_1...i_{p-1}}\delta U_{i_1...i_{p-1}} -
\dot{U}_{i_1...i_{p-1}} \delta \pi^{i_1...i_{p-1}} \right) \right\}
\nonumber\\
&-& \int_{\partial V} \left\{ (-1)^p A_{0A_2...A_p}\delta \cD^{rA_2...A_p} +
\cG^{rA_1...A_p}\delta A_{A_1...A_p} +
(-1)^p (p-1) \vp_{0A_3...A_p}\delta\rho^{rA_3...A_p} \right. \nonumber\\
  &-& \left. J^{rA_2...A_p}\delta\vp_{A_2...A_p} - (p-1)
U_{0A_3...A_p}\delta
\pi^{rA_3...A_p}
- p^{rA_2...A_p}\delta U_{A_2...A_p} \right\} \ ,
\eeq
where the ``symmetric'' Hamiltonian of the interacting electromagnetic
field and the charged matter represented by $\Phi$ reads:
\beq   \label{HM}
\lefteqn{
\cH_{sym} = \int_{V} \left\{ (-1)^p\cD^{i_1...i_p}\dot{A}_{i_1...i_p} +
(-1)^p \rho^{i_1...i_{p-1}}
\dot{\vp}_{i_1...i_{p-1}}   - \pi^{i_1...i_{p-1}} \dot{U}_{i_1...i_{p-1}}
- \cL \right.  }   \nonumber\\ &-&  \left. \p_k \left[
 (-1)^p p\, \cD^{ki_2...i_p}{A}_{0i_2...i_p} +  (-1)^p
(p-1) \rho^{ki_3...i_p}{\vp}_{0i_3...i_p} -
(p-1)  \pi^{ki_3...i_p}{U}_{0i_3...i_p}
\right]  \right\}\ .
\eeq
Now, using
\beq   \label{g}
\p_k \cD^{ki_2...i_p} = \rho^{i_2...i_p}\ ,
\eeq
one gets the following formula for $\cH_{sym}$:
\beq   \label{HM-a}
\cH_{sym} &=& \int_{V} \left\{ \frac{1}{p!} \cD^{i_1...i_p}E_{i_1...i_p}
+ (-1)^p p\, \rho^{i_1...i_{p-1}} D_0{\vp}_{i_1...i_{p-1}} -
\pi^{i_1...i_{p-1}}\dot{U}_{i_1...i_{p-1}} - \cL \right. \nonumber\\
&+& \left. \p_k\left[ (p-1)\pi^{ki_3...i_p}{U}_{0i_3...i_p} \right]
\right\}\ .
\eeq
 Moreover, due to
(\ref{g}), we may rewrite the dynamical part for $\vp$ in (\ref{4M}) as
follows:
\beq
\int_{V} \left( \dot{\rho}^{i_2...i_p}\delta\vp_{i_2...i_p} -
\dot{\vp}_{i_2...i_p}\delta\rho^{i_2...i_p} \right) &=&
\int_{V} \left( - \dot{\cD}^{ki_2...i_p} \delta (\p_k\vp_{i_2...i_p}) +
(\p_k\dot{\vp}_{i_2...i_p})\delta \cD^{ki_2...i_p} \right) \nonumber\\
&+& \int_{\p V} \left( \dot{\cD}^{rA_2...A_p} \delta\vp_{A_2...A_p}
 - \dot{\vp}_{A_2...A_p}\delta\cD^{rA_2...A_p} \right)\ .
\eeq
Now, the term  $\dot{\cD}^{rA_2...A_p}$ at the boundary may be easily
eliminated by the field equations (\ref{f-e})
\beq     \label{f???}
\dot{\cD}^{rA_2...A_p} = (-1)^p \left(  J^{rA_2...A_p} - \p_{A_1}
\cG^{rA_1A_2...A_p} \right) \ .
\eeq
Introducing  hydrodynamical variables:
\beq
V_{\mu_1\mu_2...\mu_p} :=  - D_{\mu_1} \vp_{\nu_2...\mu_p}\ ,
\eeq
we may rewrite finally (\ref{4M}) as follows:
\beq   \label{5M}
- \delta {\cH}_{sym} &=&  \int_{V} \left\{
(-1)^{p} \left( \dot{\cD}^{i_1...i_p}\delta V_{i_1...i_p}
- \dot{V}_{i_1...i_p} \delta \cD^{i_1...i_p} \right)
\right.  \nonumber\\  &-& \left.
\left( \dot{\pi}^{i_1...i_{p-1}}\delta U_{i_1...i_{p-1}} -
\dot{U}_{i_1...i_{p-1}} \delta \pi^{i_1...i_{p-1}} \right) \right\}
\nonumber\\
&-& \int_{\partial V} \left\{ (-1)^p V_{0A_2...A_p}\delta \cD^{rA_2...A_p} +
\cG^{rA_1...A_p}\delta V_{A_1...A_p} \right. \nonumber\\
  &-& \left.  (p-1)
U_{0A_3...A_p}\delta
\pi^{rA_3...A_p}
- p^{rA_2...A_p}\delta U_{A_2...A_p} \right\} \ ,
\eeq
i.e. (\ref{5M}) has exactly the same form as (\ref{4}) with $A$
replaced by the gauge-invariant $p$-form $V$ and supplemented by the
gauge-invariant canonical pair of $(p-1)$--forms $(U,\pi)$ together with
the boundary
momenta: $(p-2)$--form  $U_0$ and $(p-1)$--form $p^{r}$ on $\p V$. All
gauge-dependent terms dropped out.

\section{2 potentials vs. reduced variables}
\setcounter{equation}{0}

Let us introduce a second $p$-form gauge potential $Z$ on $\Sigma$ such that
\begin{equation}
D^{i_1...i_p} = \epsilon^{i_1...i_p k j_1...j_p} \p_k\, Z_{j_1...j_p}\ .
\end{equation}
Assuming for $Z$ the same gauge conditions as for $A$, i.e.
\beq \label{gauge1pZ}
\nb_{A_1} Z^{A_1...A_p} &=& 0\ , \\   \label{gauge2pZ}
\nb_{A_2} Z^{rA_2...A_p} &=& 0\ ,
\eeq
we have in analogy to (\ref{Ar})
\beq  \label{Zr}
\Delta_{p-1} Z^{rB_2...B_p} = (-1)^{p+1}\, \frac{r^2}{p\,p!}\,
\epsilon^{A_1...A_pB_1...B_p} \nb_{B_1} D_{A_1...A_p}\ .
\eeq
Therefore, taking into account (\ref{pi1B})--(\ref{pi2B}) one has:
\beq
\label{pi1BZ}
\Pi^1_{\ B_2...B_p} &=& (-1)^{p+1}\, \frac{r}{p}\, A_{rB_2...B_p}\ ,\\
\label{pi2BZ}
\Pi^2_{\ B_2...B_p} &=& (-1)^{p}\, \frac{r}{p}\, Z_{rB_2...B_p}\ ,
\eeq
i.e. the entire gauge-invariant information about two $p$-forms $Z$ and
$A$ on $\Sigma$ is encoded into two complex $(p-1)$--forms $Q$ and $\Pi$
on each $S^{2p}(r)$.

\section*{Acknowledgements}

This work was partially supported by the KBN Grant no 2 P03A 047 15.

\end{document}